\documentclass[structabstract]{aa}  
\usepackage{graphicx}
\usepackage{txfonts}
\usepackage{natbib}
\usepackage{longtable}
\def\pdeg{\ifmmode $\setbox0=\hbox{$^{\circ}$}\rlap{\hskip.11\wd0 .}$^{\circ}
  \else \setbox0=\hbox{$^{\circ}$}\rlap{\hskip.11\wd0 .}$^{\circ}$\fi}

\begin{document}
\title{FEEDBACK from the NGC7538 H{\sc ii} region}

\subtitle{}

   \author{H.~Beuther
          \inst{1}
          \and
          N.~Schneider
          \inst{2}
          \and
          R.~Simon
          \inst{2}
          \and
          S.~Suri
          \inst{1,3}
           \and
          V.~Ossenkopf-Okada
          \inst{2}
           \and
          S.~Kabanovic
          \inst{2}
           \and
          M.~R\"ollig
          \inst{2}
         \and
          C.~Guevara
          \inst{2}
         \and
          A.G.G.M.~Tielens
         \inst{4,5}
         \and
          G.~Sandell
         \inst{6}
         \and
          C.~Buchbender
         \inst{2}
         \and
          O.~Ricken
         \inst{7}
         \and
          R.~G\"usten
         \inst{7}
}
   \institute{$^1$ Max Planck Institute for Astronomy, K\"onigstuhl 17,
     69117 Heidelberg, Germany, \email{beuther@mpia.de}\\
     $^2$ I. Physik. Institut, University of Cologne, Z\"ulpicher Str. 77, D-50937 Cologne, Germany\\
     $^3$ University of Vienna, Department of Astrophysics, T\"urkenschanzstrasse 17, 1180 Vienna, Austria\\
     $^4$ Department of Astronomy, University of Maryland, College Park, MD 20742, USA\\
     $^5$ Leiden Observatory, Leiden University, PO Box 9513, 2300 RA Leiden, The Netherlands\\
     $^6$ Institute for Astronomy, University of Hawai`i at Manoa,  Hilo,  640 N. Aohoku Place, Hilo, HI 96720, USA\\
     $^7$ Max Planck Institute for Radioastronomy, Auf dem H\"ugel 69, 53121 Bonn, Germany 
}

   \date{Version of \today}

\abstract
    {How do expanding H{\sc ii} regions interact with their
      environmental cloud? This is one of the central questions
      driving the SOFIA legacy program FEEDBACK. Here, we present a
      case study toward the prototypical H{\sc ii} region NGC7538.}
    {We want to understand the interaction of the NGC7538 H{\sc ii}
      region with the neighboring molecular cloud that hosts 
      several active star-forming regions.}
    {With the Stratospheric Observatory for Infrared Astronomy (SOFIA)
      we mapped an area of $\sim$210$'^2$ ($\sim$125\,pc$^2$) around
      NGC7538 in the velocity-resolved ionized carbon fine-structure
      line [CII] at 1.9\,THz (158\,$\mu$m). Complementary observed
      atomic carbon [CI] at 492\,GHz and high-J CO(8--7) data as well
      as archival near-/far-infrared, cm continuum, CO(3--2) and HI
      data are folded into the analysis.}
{The ionized carbon [CII] data reveal rich morphological and kinematic
  structures. While the overall morphology follows the general ionized
  gas that is also visible in the radio continuum emission, the
  channel maps show multiple bubble-like structures with sizes on the
  order of $\sim$80--100$''$ ($\sim$1.0--1.28\,pc). While at least one
  of them may be an individual feedback bubble driven by the main
  exciting sources of the NGC7538 H{\sc ii} region (the O3 and O9
  stars IRS6 \& IRS5), the other bubble-like morphologies may also be
  due to the intrinsically porous structure of the H{\sc ii}
  region. An analysis of the expansion velocities around
  10\,km\,s$^{-1}$ indicates that thermal expansion is not sufficient
  but that wind-driving from the central O-stars is required. The
  region exhibits a general velocity gradient across, but we also
  identify several individual velocity components. The most
  blue-shifted [CII] component has barely any molecular or atomic
  counterparts.
  At the interface to the molecular cloud, we find a typical
  photon-dominated region (PDR) with a bar-shape. Ionized C$^+$,
      atomic C$^0$ and molecular carbon CO show a layered structure
  in this PDR. The carbon in the PDR is dominated by its ionized 
      C$^+$ form with atomic C$^0$ and molecular CO masses of
      $\sim$0.45$\pm0.1$\,M$_{\odot}$ and
      $\sim$1.2$\pm0.1$\,M$_{\odot}$, respectively, compared to the
      ionized carbon C$^+$ in the range of
      $3.6-9.7$\,M$_{\odot}$.. This bar-shaped PDR exhibits a
  velocity-gradient across, indicating motions along the
  line-of-sight towards the observer.}
{Even if dominated by two nearby exciting sources (IRS6 \& IRS5), the
  NGC7538 H{\sc ii} region exhibits a diverse set of sub-structures
  that interact with each other as well as with the adjacent
  cloud. Compared to other recent [CII] observations of H{\sc ii}
  regions (e.g., Orion Veil, RCW120, RCW49), bubble-shape morphologies
  revealed in [CII] emission, indicative of expanding shells, are
  recurring structures of PDRs.}


\titlerunning{FEEDBACK from the NGC7538 H{\sc ii} region}

\maketitle
 
\section{Introduction}
\label{intro}

Stellar feedback in the form of radiation, winds and supernova
explosions can have positive as well as negative impact on their
environment. Positive feedback here is defined as inducing new star
formation processes whereas negative feedback is meant as a
destruction process of the natal cloud and by that limiting further
star formation. Feedback is important for the determination of the
physical processes within individual clouds but also crucial in an
integral sense for whole galaxies because it likely determines the
general star formation efficiency (e.g.,
\citealt{matzner2002,elmegreen2011b,hopkins2014,geen2016,geen2018,kim2018}).

With the goal to understand the impact of feedback processes onto the
environment for individual resolved cases, the SOFIA (Stratospheric
Observatory for Infrared Astronomy) legacy program
FEEDBACK\footnote{http://feedback.astro.umd.edu} targets 11 galactic
high-mass star-forming regions and H{\sc ii} regions in the ionized
atomic carbon fine-structure line [CII] at 158\,$\mu$m (1.9\,THz) and
the atomic oxygen line [OI] at 63\,$\mu$m (4.7\,THz). The ionized
carbon line [CII] is of particular importance because it allows us to
study the kinematics of the ionized gas as well as the interfaces with
the molecular clouds. The [CII] line is a direct tracer of
bubble/stellar feedback kinematics (e.g.,
\citealt{pabst2020}). Furthermore, the [CII] line is one of the
dominant cooling lines in the interstellar medium at low to
intermediate densities and UV fields (e.g.,
\citealt{hollenbach1997,roellig2013}).  The FEEDBACK program started
in spring 2019 and is still ongoing. More details about this SOFIA
legacy program are provided in \citet{schneider2020}. One outcome of
the FEEDBACK project so far is the detection of expanding shells in
[CII] emission in various bubble-shaped or bipolar H{\sc ii} regions,
i.e., in RCW120 (\citealt{luisi2021}, Kabanovic et al.~in press),
RCW49 \citep{tiwari2021}, RCW36 (Bonne et al., subm.), and RCW79
(Zavagno et al., in prep.).  These shells were first detected in Orion
A \citep{pabst2019} and appear to be a ubiquitous phenomena. While the
regions RCW49 and RCW120 reveal second generation star formation
within these shells \citep{tiwari2021,luisi2021}, the expanding veil
nebula in Orion shows little evidence for dense clumps that could lead
to star formation \citep{pabst2019,goicoechea2020}.

\begin{figure}[htb]
\includegraphics[width=0.49\textwidth]{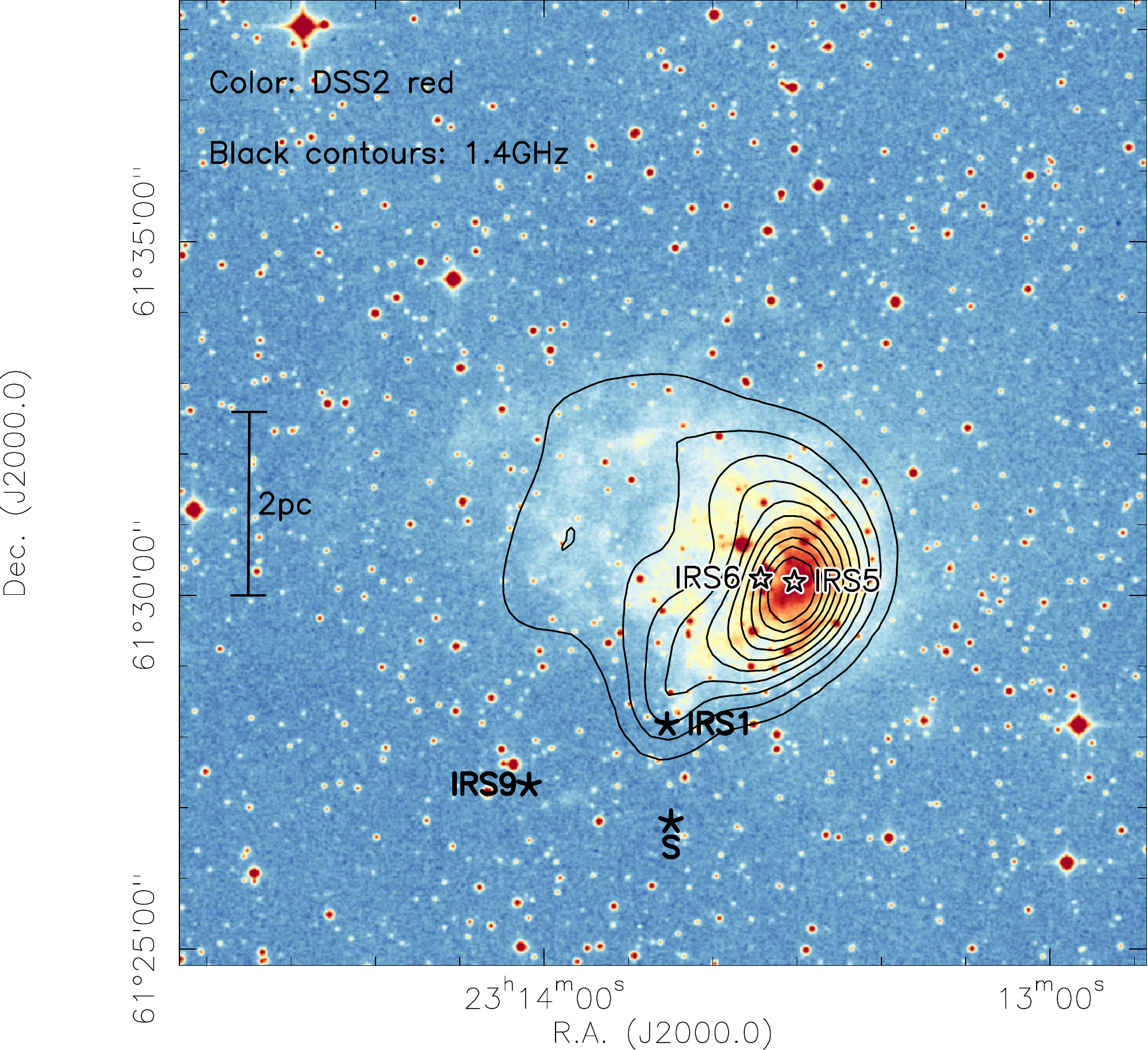}
\caption{Optical and cm continuum data of the NGC7538 region,
  encompassing the field of view of our observations. The color-scale
  shows the optical DSS2 (Digital Sky Survey v2) obtained with the red
  filter.  The contours outline the 1.4\,GHz cm continuum emission at
  $1'$ angular resolution from the Canadian Galactic Plane Survey
  (CGPS, \citealt{taylor2003}). The two stars in the center of the
  H{\sc ii} region mark the positions of IRS5 and IRS6 from
  \citet{puga2010}. The three 5-pointed stars directly below the H{\sc
    ii} regions show the positions of the active sites of star
  formation IRS1, S and IRS9.}
\label{dss2} 
\end{figure}

While an analysis of the whole sample will follow when all data are
taken, in the following, we concentrate on the prototypical H{\sc ii}
region NGC7538 shown at optical and cm continuum wavelengths in Figure
\ref{dss2}. The region is located at a distance of 2.65\,kpc
\citep{moscadelli2009} and has a physical extent of several
parsec. The two brightest sources exciting the H{\sc ii} region are
IRS5 and IRS6 with spectral types of O9 and O3, respectively
(\citealt{puga2010}, marked in Fig.~\ref{dss2}). The H{\sc ii} region
is associated with several sites of active star formation, in
particular in the south with the regions IRS1, S and IRS9 (marked in
Fig.~\ref{dss2}). Towards these southern star-forming regions, the
H{\sc ii} region appears to be sharply bounded, whereas in the
northeast the region shows diffuse emission extending well beyond the
photon-dominated region (PDR) shell \citep{luisi2016}. In a recent
study combining CO(3--2) and [CII] data, \citet{sandell2020}
characterized a large north-south outflow emanating from the southern
star-forming core IRS1 with blue-shifted ionized carbon [CII]
and molecular CO emission north of IRS1 in the vicinity of the H{\sc
  ii} region.  Furthermore, \citet{townsley2018} show that diffuse
X-ray emission is found toward the H{\sc ii} region, spreading even
beyond the extend of the H{\sc ii} region as determined by, e.g., the
cm continuum emission.

\begin{figure*}[htb]
\includegraphics[width=0.99\textwidth]{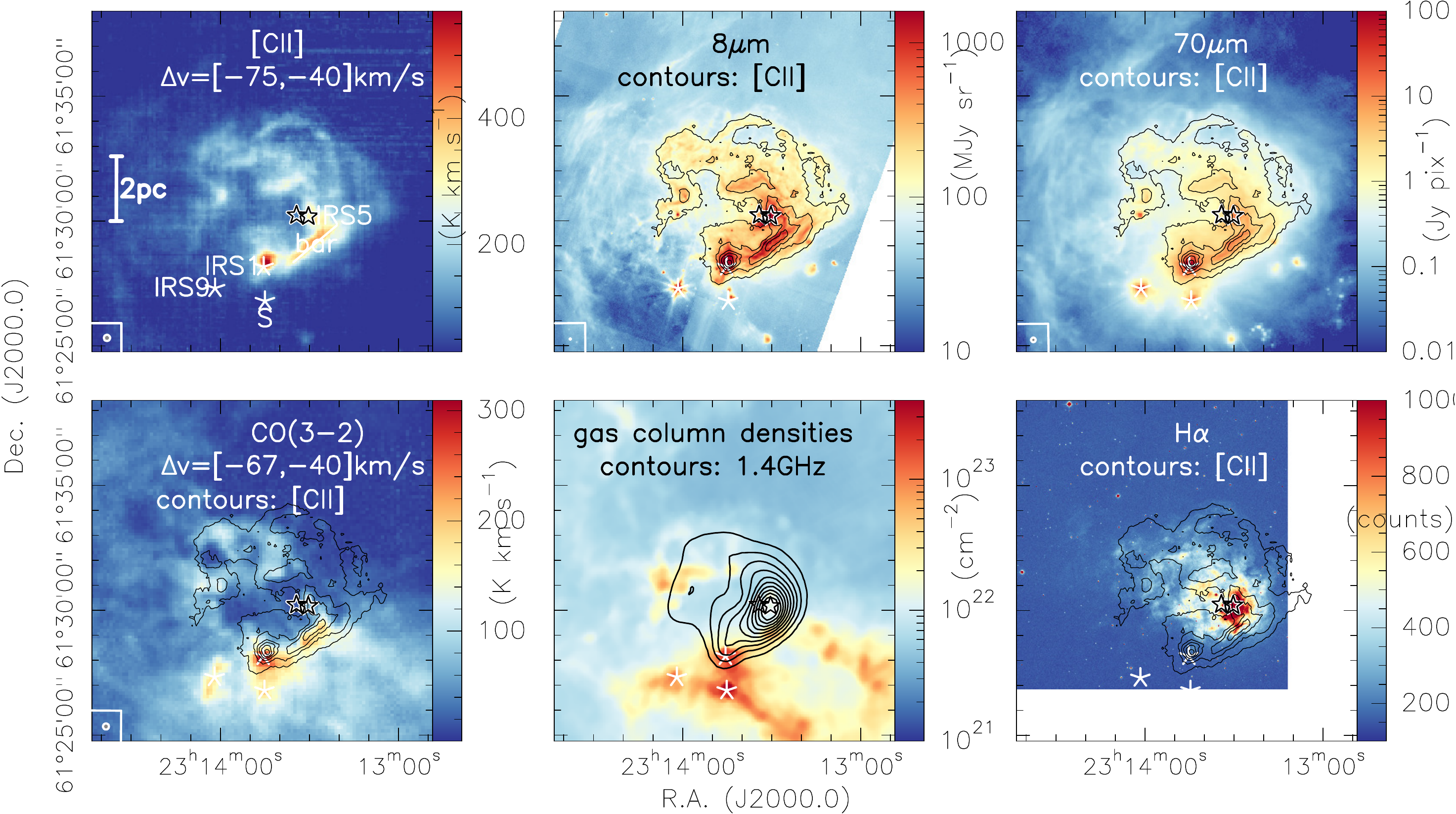}
\caption{Integrated intensity molecular and atomic line maps as well
  as continuum emission are presented in color-scale toward
  NGC7538. The specific lines, integration ranges and continuum bands
  are labeled in each panel. The corresponding angular
  resolution is shown at the bottom-left of the panels. The contours
  in most panels as labeled show the [CII] emission at only a few
  selected contour levels to highlight similarities and
  differences. The contours in the bottom-middle panel outline the
  1.4\,GHz cm continuum emission from the CGPS. The three five-pointed
  stars mark the positions of the embedded sources NGC7538IRS1,
  NGC7538S and NGC7538IRS9. The two stars in the middle show the
  positions of the two main exciting sources of the H{\sc ii} region
  NGC7538IRS5 and NGC7538IRS6. The bar-like PDR as well as a a linear
  scale-bar are presented in the top-left panel.}
\label{mom0} 
\end{figure*}

More generally speaking, the NGC7538 H{\sc ii} region is embedded in a
larger molecular cloud complex (e.g.,
\citealt{ungerechts2000,fallscheer2013}). Figure \ref{mom0} gives an
overview of various observed tracers. The cm continuum and 8$\mu$m
emission appears to stem dominantly from the H{\sc ii} region. One
should keep in mind that $8\mu$m emission is usually emitted in the
cavity walls of the H{\sc ii} region where the poly-cyclic aromatic
hydrocarbons (PAH) survive. In contrast to these features, the dense
gas as traced by the Herschel dust continuum or CO(3--2) emission is
found more in the outskirts of the H{\sc ii} region, in particular
towards the south where the well-know star-forming regions IRS1, S and
IRS9 are located. As will be discussed in more detail below, the new
SOFIA [CII] data combine different worlds: They trace the PDR, and
they also show strong emission features toward the active star-forming
region IRS1 and a bar-like region at the south-eastern interface of
the H{\sc ii} region with the dense molecular cloud. Hence, the
velocity-resolved [CII] data are the ideal probe to study the feedback
processes and the kinematic gas properties from the evolving H{\sc ii}
region on its neighboring molecular cloud.

Specific topics we address with this study are the impact from the
expanding H{\sc ii} region onto the environmental dense gas. Can we
identify expanding [CII] shells or other dynamic features, either
triggering or preventing new star formation processes? Do we identify
layered structures from a photon-dominated region (PDR)? What is the
carbon budget between ionized C$^+$, atomic C$^0$ and molecular CO gas
specifically at the H{\sc ii} region/molecular cloud interface?

\section{Observations and data}
\label{data}

\subsection{SOFIA observations}

The region was mapped with the dual-frequency array heterodyne
receiver upGREAT\footnote{German Receiver for Astronomy at
  Terahertz. (up)GREAT is a development by the MPI for Radioastronomy
  and the KOSMA/University of Cologne, in cooperation with the DLR
  Institute for Sensorsystems.} \citep{risacher2018} in the
velocity-resolved [CII] fine-structure line at
158\,$\mu$m/1.9\,THz\footnote{The CII 158\,$\mu$m data are provided at
  the IRSA/IPAC Infrared science archive
  https://irsa.ipac.caltech.edu/Missions/sofia.html.}. The [OI]
63\,$\mu$m line was observed in parallel.  Fast Fourier Transform
Spectrometers (FFTS) with 4\,GHz instantaneous bandwidth
\citep{klein2012} resulted in a nominal spectral resolution for the
      [CII] line of 0.244\,MHz or 0.04\,km\,s$^{-1}$. The observations
      were done on two consecutive days during Cycle 7, on December
      12th and 13th 2019, with an average $T_{\rm sys}$ single
      side-band (SSB) of 4486 and 4580~K and an average precipitable
      water vapor of 6.2 and 3.6~$\mu$m respectively per day. The data
      were calibrated to the main beam brightness temperature
      intensity scale, $T_{\rm mb}$, with the {\it kalibrate} task
      \citep{guan2012}, part of the standard GREAT pipeline, with a
      forward efficiency of 0.97 and an average beam efficiency of
      0.66.

Mapping was conducted in an on-the-fly (OTF) mode where the entire
NGC7538 map was split into 4 tiles of roughly $436''$ on each
side. During scanning the data were better than Nyquist-sampled with a
dump at every $5.2''$. The final map size is $14.5'\times 14.5' =
11.2\rm{pc} \times 11.2\rm{pc}$. Each tile should typically be
observed 4 times with different scanning angles to reduce potential
striping effects. In the case of NGC7538, three tiles were mapped to a
deeper sensitivity whereas for the fourth north-western tile not all
coverages were conducted so far. This results in slightly increased
noise in the top-right tile of the map and residual striping quite
apparent in Fig.~\ref{mom0}. The data cubes used here have been
resampled at a velocity resolution of 0.5\,km\,s$^{-1}$ to increase
the signal-to-noise ratio. Our final velocity-resampled and grided
data-cube has an angular resolution of $15.8''$ (corresponding
to $\sim$0.2\,pc linear spatial resolution). The $1\sigma$ rms
over the well-covered 3/4 of the map are $\sim$0.9\,K per
0.5\,km\,s$^{-1}$ channel on a $T_{\rm mb}$ scale. The higher
$1\sigma$ rms in the north-western tile is $\sim$1.9\,K per
0.5\,km\,s$^{-1}$ channel.

Furthermore, slightly smaller maps of the region were observed in the
atomic carbon fine-structure line [CI] at 492\,GHz and three CO
transitions with $J=8-7$, $J=11-10$ and $J=22-21$. Here, we are
concentrating on the lower-density tracer [CI] as well as the CO(8--7)
transition at 921.7997\,GHz.  The [OI] 63\,$\mu$m and higher-J CO
lines, tracing higher densities and temperatures, will be analyzed in
forthcoming studies. The angular resolution of the final data cubes
for the [CI] and CO(8--7) lines are $63''$ and $42''$,
respectively. The $1\sigma$ rms values for the [CI] and CO(8--7) lines
at 0.5\,km\,s$^{-1}$ are 0.7 and 1.4\,K, respectively.

\subsection{Complementary archival data}

\subsubsection{Optical, mid-infrared, far-infrared and radio data}

We used the Digital Sky Survey version 2 (DSS2) to obtain an optical
image of the region in the red filter. The DSS2 digital images are
based on photographic images obtained using the Oschin Schmidt
Telescope of Palomar Mountain and the UK Schmidt Telescope. The plates
were processed into the present compressed digital form with the
permission of the institutions.

The 1.4\,GHz radio continuum and the 21\,cm HI data are from the
Canadian Galactic Plane Survey (CGPS, \citealt{taylor2003}). The
angular resolution of these data products is $\sim 1'$. The $1\sigma$
rms of the continuum and line data are $\sim 0.5$\,K and 3\,K in
0.82\,km\,s$^{-1}$ channels, respectively.

The 8\,$\mu$m photometric emission mid-infrared data are obtained with
the IRAC camera \citep{fazio2004} on board of the Spitzer satellite
\citep{werner2004} at an angular resolution of $\sim 2''$.

The 70\,$\mu$m far-infrared data are observed with the PACS camera
\citep{A&ASpecialIssue-PACS} on board of the Herschel satellite
mission \citep{A&ASpecialIssue-HERSCHEL} in the framework of the HOBYS
key program \citep{motte2010}. The angular resolution is $\sim 5.6''$.

The H$_{\alpha}$ data were observed in the framework of ``The INT
Photometric H$_{\alpha}$ Survey of the Northern Galactic Plane''
(IPHAS) at an angular resolution of $\sim 1.1''$ \citep{drew2005}.

\begin{figure*}[htb]
\includegraphics[width=0.99\textwidth]{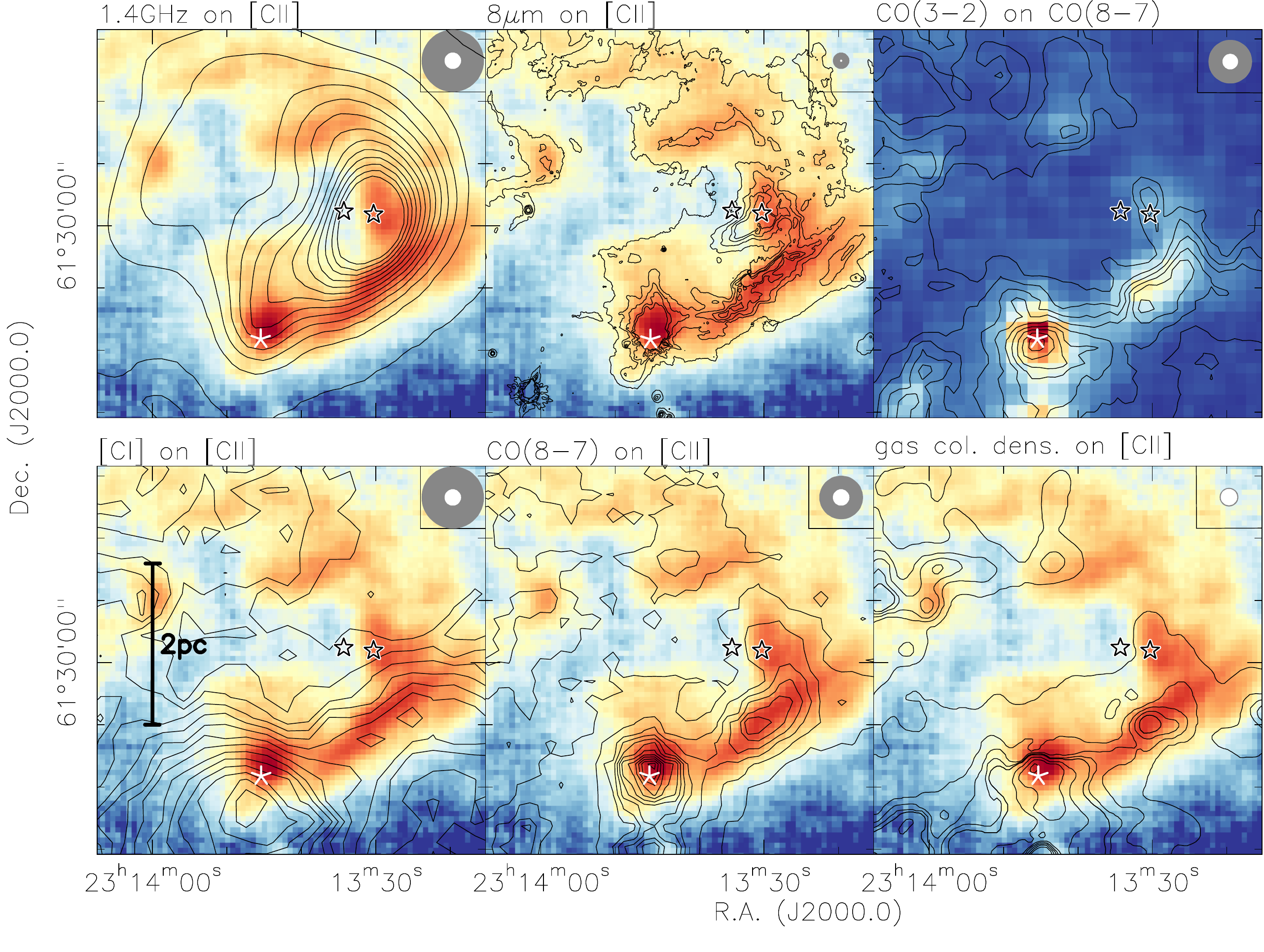}
\caption{Overlays of selected pairs of emission tracers as described
  above each panel. The integration ranges for the atomic and
  molecular lines are the same as in Fig.~\ref{mom0}. The contour
  levels are chosen individually for each figure to highlight the
  emission best. The angular resolution elements for both tracers are
  presented in the top-right of each panel. The five-pointed stars
  always mark the position of the embedded source NGC7538IRS1, and the
  two open stars show the positions of the two main exciting sources of the
  H{\sc ii} region NGC7538IRS5 and NGC7538IRS6. A linear scale-bar is
  presented in the bottom-left panel.}
\label{overlays} 
\end{figure*} 

\subsubsection{Herschel column density and CO(3--2) data}

The Herschel gas column density map we use in this paper is a higher
angular resolution version (18$''$) of the one presented in
\citet{fallscheer2013}.  The map was produced by the HOBYS (Herschel
imaging survey of OB Young Stellar objects, \citealt{motte2010})
consortium and will soon be publicly provided. The column density map
was determined following the HOBYS procedure, i.e., from a
pixel-to-pixel spectral energy distribution (SED) greybody fit to the
160, 250, 350, and 500\,$\mu$m wavelengths observations.  There were
no 100\,$\mu$m maps available and the 70\,$\mu$m data are not included
because the interest lies in the distribution of the cold molecular
gas. For the SED fit, we fixed the specific dust opacity per unit mass
(dust+gas) approximated by the power law $\kappa_\nu=0.1
(\nu/1000\rm{GHz})^{\beta_d} {\rm cm}^2/{\rm g}$ with $\beta_d$=2, and
left the dust temperature and column density (atomic and molecular
gas, in the following we refer to that as gas column density) as free
parameters (see, e.g., \citealt{russeil2013} for more details). The
procedure explaining how high angular resolution maps were obtained is
described in detail in Appendix A of \citet{palmeirim2013}. The
concept is to employ a multi-scale decomposition of the flux maps and
assume a constant line-of-sight temperature. The final map at 18$''$
angular resolution is constructed from the difference maps of the
convolved SPIRE maps (at 500, 350, and 250 $\mu$m) and the temperature
information from the color temperature derived from the 160 to 250
$\mu$m ratio.

The CO(3--2) data were obtained with the James Clark Maxwell Telescope
(JCMT) as part of a large map of the NGC7538 region first presented in
\citet{fallscheer2013} and later also in
\citet{sandell2020}. Following \citet{fallscheer2013}, the typical
$1\sigma$ rms in 0.42\,km\,s$^{-1}$ channels is 0.6\,K on a $T_A^*$
scale. Using a beam efficiency of
0.64\footnote{https://www.eaobservatory.org/jcmt/instrumentation/heterodyne/harp/},
the data are converted to main beam brightness temperature for the
quantitative analysis. The beam size of these data is $\sim 15''$.

\begin{figure*}[htb]
\includegraphics[width=0.99\textwidth]{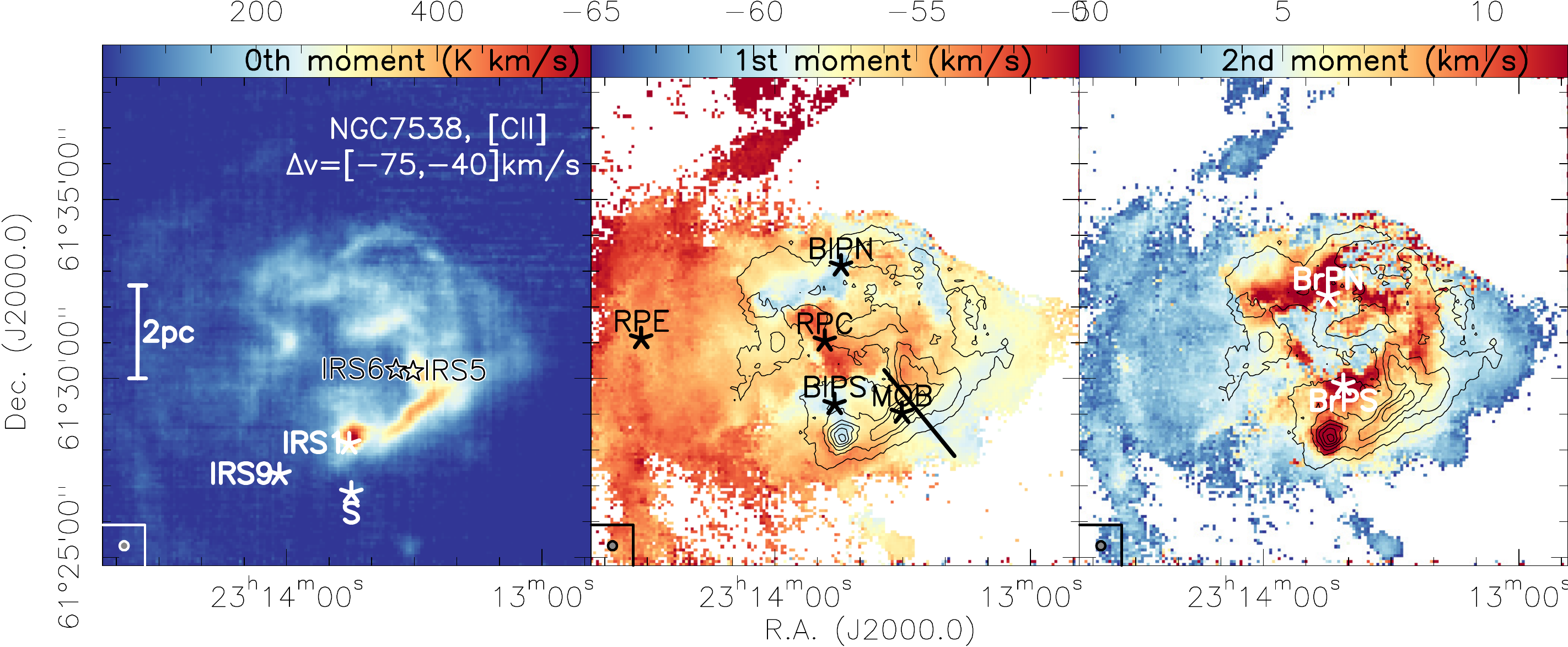}
\caption{Moment maps of the [CII] emission. The left, middle and right
  panels present in color scale the integrated emission and the 1st
  and 2nd moment maps (intensity-weighted peak velocities and
  intensity-weighted velocity dispersion) in the velocity range from
  -75 to -40\,km\,s$^{-1}$, respectively. The noisier top-right
  quarters of the 1st and 2nd moment maps are blanked (see
  Sect.~\ref{data}).  The contours in the middle and right panels show
  the integrated [CII] emission from the left panel at only a few
  selected contour levels to highlight similarities and differences.
  The three five-pointed stars in the left panel mark the positions of
  the embedded sources NGC7538IRS1, NGC7538S and NGC7538IRS9. The two
  stars show the positions of the two main exciting sources of the
  H{\sc ii} region NGC7538IRS5 and NGC7538IRS6. A linear scale-bar is
  presented in the left panel. The markers in the middle and right
  panels highlight additional positions (see main text) toward where
  we extracted spectra in Fig.~\ref{spectra}. The line in the middle
  panel outlines the pv-cut from IRS6 through the bar-like structure
  presented in Fig.~\ref{pv_bar}.}
\label{moments} 
\end{figure*}

\section{Results}
\label{results}

\subsection{Spatial structures}

As indicated in the Introduction, the [CII] emission in the NGC7538
complex traces the large-scale PDR as well as the PDR associated with
denser star-forming molecular gas at the edge of the H{\sc ii}
region. To be more specific, Fig.~\ref{mom0} shows the integrated
emission of several tracers, and in particular the morphology of the
[CII], the 8\,$\mu$m and 70\,$\mu$m emission is very similar. The
ionized gas as traced by the H$_{\alpha}$ and 1.4\,GHz emission is
confined to the real H{\sc ii} region, and the [CII], 8\,$\mu$m and
70\,$\mu$m emission partly wrap around that (bottom-right panel of
Fig.~\ref{mom0}). While the [CII] line can form in PDRs as well as
molecular media, diffuse 8\,$\mu$m emission typically stems from
UV-pumped infrared fluorescence from Polycyclic Aromatic Hydrocarbon
(PAH) molecules within photon-dominated regions (PDRs). The dense gas
tracers (gas column densities from dust continuum and CO(3--2), bottom
panels in Fig.~\ref{mom0}) emit dominantly in the south and south-west
of the region. We note that toward the largest part of the [CII]
emission the CO(3--2) line does not show significant self-absorption
features (see spectra discussed below). The associated emission toward
the young embedded star-forming region IRS1 as well as the bar-like
PDR features are visible in all tracers shown in Fig.~\ref{mom0}. In
addition to this, some dense gas structures can also be found in the
north-eastern part of NGC7538.

To show spatial similarities and differences in a bit more detail,
Figure \ref{overlays} presents a compilation of different pairs of
tracers always as contour maps overlaid on color-maps. We added here
the higher excited CO(8--7) line ($E_u/k=199$\,K compared to
$E_u/k=33.2$\,K for the (3--2) transition) as well as the atomic
carbon fine structure line [CI] at 492\,GHz. The top-left and
top-middle panels confirm that the [CII] and 8\,$\mu$m emission agree
very well and appear to at least partially wrap around the ionized gas
(see also bottom-right panel of Fig.~\ref{mom0}).  A comparison of the
two CO transitions in the top-right panel shows that in the
south-western bar-like emission structure, the higher excited CO(8-7)
transition emits closer to the two exciting sources at the center of
the H{\sc ii} region than the lower excited CO(3--2)
transition. Similar layered structures within the bar can be
identified between atomic and ionized carbon ([CI] and [CII],
bottom-left panel of Fig.~\ref{overlays}) as well as between the
CO(8--7) and [CI] maps. In comparison to those layered structures,
spatial morphologies of the higher excited CO(8--7) and ionized carbon
[CII] lines agree much better in the bar. This is a typical PDR
layered structure in [CII]/[CI]/CO as well as in CO(8-7) and (3-2). We
will come back to this in section \ref{pdr}.

\begin{figure*}[htb]
\includegraphics[width=0.99\textwidth]{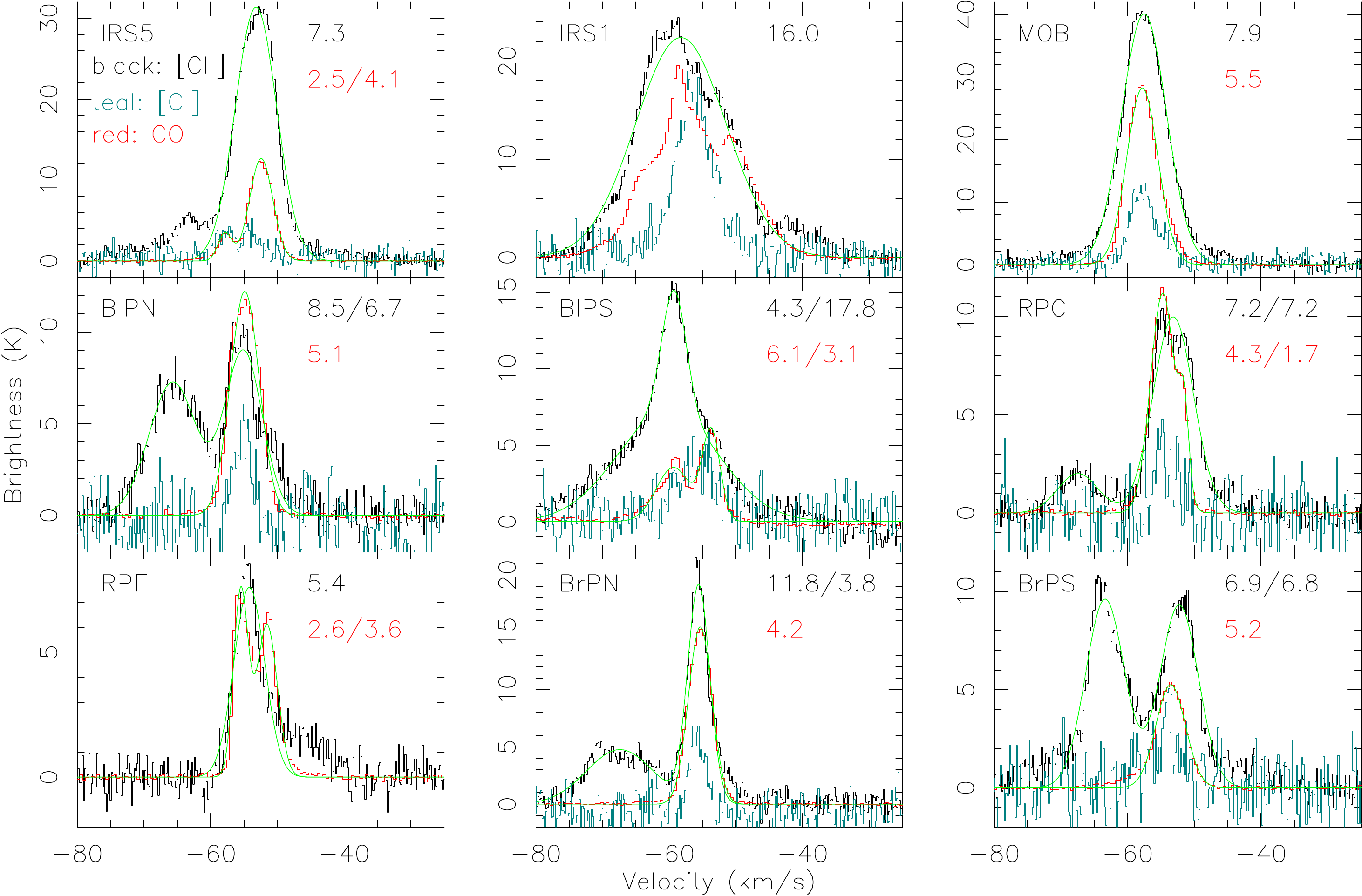}
\caption{[CII] (black), CO(3--2) (red) and [CI] (teal, multiplied by
  2) spectra towards the positions marked in Fig.~\ref{moments}. The
  green lines show Gaussian fits to the [CII] and CO(3--2) data. The
  numbers present the corresponding FWHM values (``*/*'' labels two
  components). The positions are labeled in each panel. The RPE
  position to the east is not covered by the [CI] map.}
\label{spectra} 
\end{figure*}

\subsection{Kinematic structures}
\label{kinematics}

To get a first proxy of the dynamics, Fig.~\ref{moments} presents the
integrated [CII] emission as well as the 1st and 2nd moment maps of
the [CII] data (intensity-weighted peak velocities and velocity
dispersion).  The 1st moment map (middle panel of Fig.~\ref{moments})
reveals a general velocity gradient approximately from east to west
over the entire extend of the H{\sc ii} region. On top of the overall
velocity gradient, one can identify several blue-shifted emission
regions in the middle of the map. In comparison to that, the 2nd
moment or velocity dispersion map shows over large parts of the map
rather uniformly low values below 5\,km\,s$^{-1}$. However, on top of
that, one can clearly identify an almost ring-like structure with
larger velocity dispersion on the order of 10\,km\,s$^{-1}$. As will
be discussed below, these regions of apparent large velocity
dispersion are largely caused by multiple velocity components in the
[CII] spectra.

For a more detailed analysis of the kinematics, it is crucial to look
at individual spectra, and we selected several positions in the region
for further analysis. We labeled them as follows, and they are all
marked in Fig.~\ref{moments}:\\

\noindent - IRS5: one of the main exciting source of the H{\sc ii} region\\
- IRS1: the densest star-forming core\\
- MOB: a position in the middle of the bar-like PDR\\
- BlPN: blue peak in the north in the 1st moment map\\
- BlPS: blue peak in the south of the 1st moment map\\
- RPC: red peak in the center of the 1st moment map\\
- RPE: red peak in the east of the 1st moment map\\
- BrPN: broad emission in the north in the 2nd moment map\\
- BrPS: broad emission in the south in the 2nd moment map\\

Figure \ref{spectra} presents the corresponding [CII] spectra as well
as the molecular and atomic counterparts in the CO(3--2) and [CI]
lines\footnote{Fig.~\ref{spectra_grid} in the Appendix presents a more
  extended grid of spectra in the [CII] and CO(3--2) lines.}. Where
possible we fitted Gaussians to the [CII] and CO(3--2) spectra, at
several positions also two components. While self-absorption may
explain smaller dips in some spectra, the two velocity components
fitted here are typically that far separated in velocity space that
they should indeed be separate components. The corresponding full
width half maximum values (FWHM) in km\,s$^{-1}$ are also shown in
Fig.~\ref{spectra}. Interestingly, the only almost simple,
single-Gaussian spectrum is identified toward the position in the
middle of the bar (MOB, top-right panel in Fig.~\ref{spectra}). We
will get back to that structure in section \ref{pdr}. The spectra
toward the infrared sources IRS5 and IRS1 as well as the red-shifted
peak in the east of the region (RPE) show some velocity structure, but
no particularly prominent features.

This is very different in the remaining five spectra towards the other
blue- and red-shifted positions in the center of the maps (BlPN, BlPS,
RPC) as well as towards the positions with particularly broad velocity
dispersion (BrPN, BrPS, Figs.~\ref{moments} and \ref{spectra}). While
the main emission from the NGC7538 region is typically in the velocity
range with respect to the local standard of rest ($v_{\rm LSR}$)
between -60 and -50\,km\,s$^{-1}$, these positions show additional
strong emission at even more negative velocities beyond
-60\,km\,s$^{-1}$. In particular, the spectra toward BlPN, RPC, BrPN
and BRPS exhibit a second, clearly separated velocity component in the
[CII] emission that has barely any detectable counterpart in the
molecular emission of the CO(3--2) line or the atomic [CI]
emission. The rms values for these two lines (section \ref{data})
correspond at 30\,K to $3\sigma$ column density sensitivities (see
section \ref{carbon_budget} for details on the calculations) of
$N(\rm{CO})\approx 2.1\times 10^{16}$\,cm$^{-2}$ and
$N(\rm{C^0})\approx 2.7\times 10^{16}$\,cm$^{-2}$ per channel (0.42
and 0.5\,km\,s$^{-1}$ for CO(3--2) and [CI], respectively). This
second component can also be identified towards IRS5 and BlPS. Hence,
the ring-like high velocity dispersion structure seen in the 2nd
moment map of [CII] (Fig.~\ref{moments}, right panel) is in fact not a
region of particularly high velocity dispersion but there we have two
velocity components mimicking high values in the 2nd moment map. The
peculiar aspect of this additional velocity component is that it is
mainly detected in the ionized carbon [CII] line without a strong
molecular or atomic counterpart being detectable in the individual
spectra. However, \citet{sandell2020} detected a north-south outflow
in CO(3--2) and [CII] emission (their [CII] map is centered on IRS1
and smaller than the one presented here) emanating from IRS1 where the
blue-shifted emission is located north of IRS1. The positions BlPN,
BlPS, RPC, BrPN, and BrPS are all in the general vicinity of that
outflow. In particular, the morphology of the blue-shifted [CII] and
CO(3-2) gas emission near IRS1 (corresponding roughly to our position
BlPS) is similar. Hence, the blue-shifted gas we find may at least
partly be associated with that large-scale outflow. Nevertheless, in
the spectra extracted at individual positions as shown in Figure
\ref{spectra}, the blue-shifted gas is much stronger in the [CII]
spectra than in the more commonly as outflow tracer observed CO
emission. We will get back to this component in Section
\ref{decoupled}.

\begin{figure*}[htb]
\includegraphics[width=0.99\textwidth]{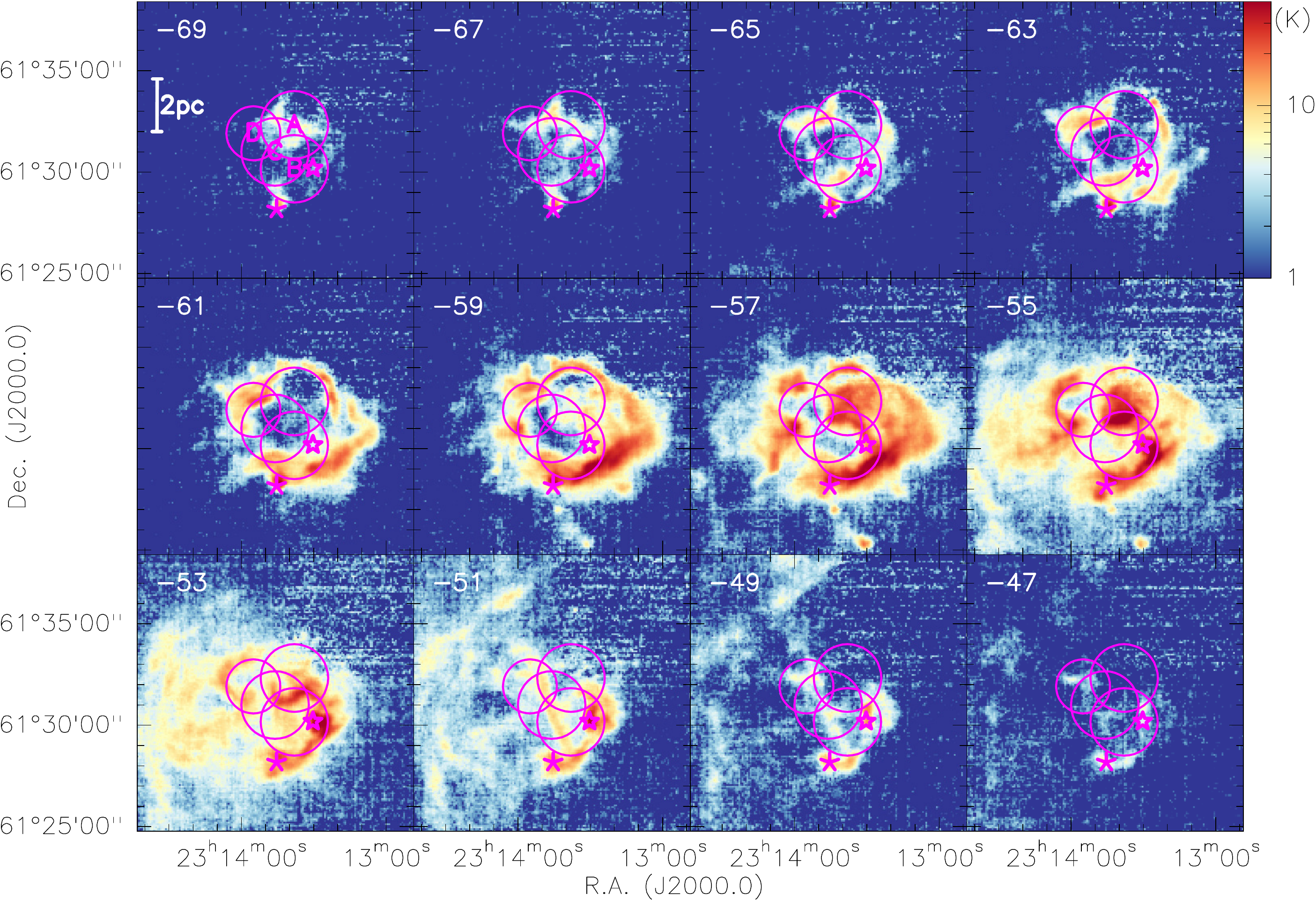}
\caption{[CII] channel map for the velocities labeled in each
  panel. The data are binned for that plot in 2\,km\,s$^{-1}$
  channels. The two purple stars mark the positions of IRS1 and IRS5,
  and a scale-bar is shown in the top-left panel. The intensities are
  always plotted logarithmically from 1 to 40\,K. The purple circles
  outline potential bubbles. These bubbles are labeled in the
  top-left panel. A version without any suggested bubble structures
  is presented in the Appendix in Fig.~\ref{channel_cii_nobubbles}.}
\label{channel_cii} 
\end{figure*} 

To look at the velocity structures in a different fashion,
Fig.~\ref{channel_cii} presents a channel map of the [CII] emission in
NGC7538 (a version without any suggested bubble structures is
presented in the Appendix in Fig.~\ref{channel_cii_nobubbles}).  While
in this representation the main emission of the PDR associated with
the H{\sc ii} regions is also dominant approximately between -60 and
-50\,km\,s$^{-1}$, there are considerable extended emission structures
at blue-shifted velocities $<-60$\,km\,s$^{-1}$, and to a lesser
degree also at red-shifted velocities $>-50$\,km\,s$^{-1}$. For
comparison, Figure \ref{channel_co} in the appendix presents the
corresponding CO(3--2) channel map where especially these more blue-
and more red-shifted features are barely recognizable.  The [CII]
channel map gives the impression of several bubble- and ring-like
structures that may stem even from several star formation events. We
will discuss the bubble-features and their possible interpretation in
more detail below in Section \ref{bubbles}.

In addition to the molecular CO, atomic [CI] and ionized carbon [CII]
emission, we can also investigate the atomic hydrogen by means of the
21\,cm line observed with the Canadian Galactic Plane Survey (CGPS,
\citealt{taylor2003}). As expected, towards the main H{\sc ii} region
around the exciting sources IRS5 and IRS6, the HI is seen only in
absorption. One clearly identifies a strong HI absorption component
associated with the high-velocity blue-shifted gas
$<-60$\,km\,s$^{-1}$, but the main velocity component of the region
around $-55$\,km\,s$^{-1}$ exhibits only weak absorption. The
blue-shifted HI absorption should be related to the expanding shell
accelerating also the atomic gas envelope in the direction of the
observer. The red-shifted absorption at $\sim-45$\,km\,s$^{-1}$ cannot
be the other side of the expanding shell because in absorption
spectroscopy it has to lie in front of the ionized gas. Hence, the
red-shifted HI absorption must be related to some foreground
gas. Since there is barely any [CII] nor CO emission at these
velocities, that component may even be unrelated to the NGC7538 H{\sc
  ii} region.

\begin{figure}[htb]
\includegraphics[width=0.49\textwidth]{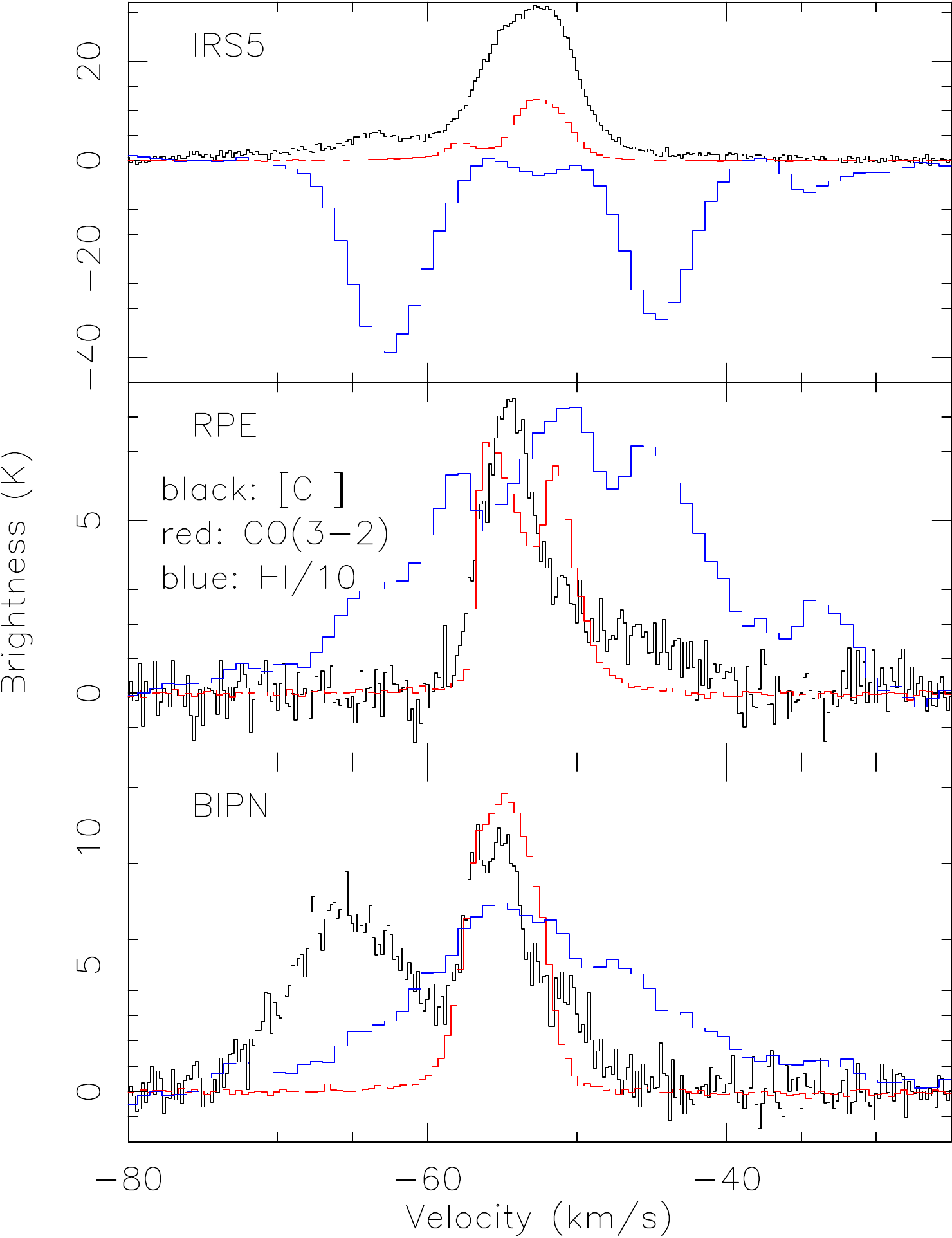}
\caption{[CII] (black), CO(3--2) (red) and HI (blue, divided by 10)
  spectra towards three selected positions marked in
  Fig.~\ref{moments}. These are IRS5, the red-shifted peak towards the
  east (RPE), and the blue-shifted peak in the north (BlPN).  The
  positions are labeled in each panel. }
\label{spectra_hi} 
\end{figure}

Looking a bit outside the actual H{\sc ii} region towards the east
(position RPE, middle panel of Fig.~\ref{spectra_hi}), the HI emission
spectrum is broader than what we observe in [CII] and CO. This can be
understood in a way that HI is more easily excited than [CII] and CO,
and hence can pick up more tenuous gas at higher
velocities. Furthermore, at the velocity of the [CII] and CO(3--2)
peak emission, the HI shows a dip in the spectrum which may be caused
by HI self-absorption (HISA, e.g., \citealt{gibson2005b,syed2020}).

Furthermore, if we look towards positions where we clearly identify
two components in the [CII] emission, e.g, the position BlPN in the
north of the H{\sc ii} region (bottom panel of Fig.~\ref{spectra_hi}),
it may at first sight be surprising that the strong blue-shifted [CII]
component at velocities $<-60$\,km\,s$^{^1}$ is inconspicuously weak
in the HI emission. One potential explanation for that could be that
the radiation field may be that strong that most of the gas is ionized
hydrogen and carbon with just a thin layer of neutral HI and molecular
gas in the surrounding molecular cloud.

\section{Discussion}
\label{discussion}

\subsection{Expanding bubbles or an inhomogeneous medium?}
\label{bubbles}

\paragraph{Bubble/ring identification:} As indicated in section \ref{kinematics}, the channel map of the [CII]
emission shows several ring-like structures by eye-inspection (four
rings are outlined in \ref{channel_cii} as magenta circles). These
features can be limb-brightened edges of expanding shells or bubbles,
but they could also be a ring or torus in the plane of the sky. Before
we discuss their velocity structure in more detail, we apply an
unbiased automized ring identification by computing the covariance
between the map and annuli structures of varying radii. The details of
the identification analysis are presented in appendix
\ref{bubble_finding}.W e stress that the bubbles/rings discussed below
are only candidates for real physical structures that may stem from
the expansion of the H{\sc ii} region. The bubble- or ring-like
structures we identify this way have central positions (in J2000.0) and approximate
radii $r$ of:\\ (A) R.A.=23h13m38s,
Dec=61$^{\circ}$32$'$20$''$; $r\sim 100''\sim$1.3\,pc\\ (B)
R.A.=23h13m38s, Dec=61$^{\circ}$30$'$12$''$; $r\sim 100''\sim$1.3\,pc\\ (C) R.A.=23h13m46s, Dec=61$^{\circ}$31$'$00$''$; $r\sim
100''\sim$1.3\,pc \\ (D) R.A.=23h13m55s, Dec=61$^{\circ}$31$'$56$''$;
$r\sim 80''\sim$1.0\,pc  \\ As outlined in appendix \ref{bubble_finding}, the
approach identifies even a fifth ring-like structure (labeled E),
however, that appears rather as an overlap of mainly structures A to
C, and we do not consider that further as a separate physical
entity. Uncertainties for the central positions and radii are also
discussed in appendix \ref{bubble_finding}.

\begin{figure}[htb]
\includegraphics[width=0.49\textwidth]{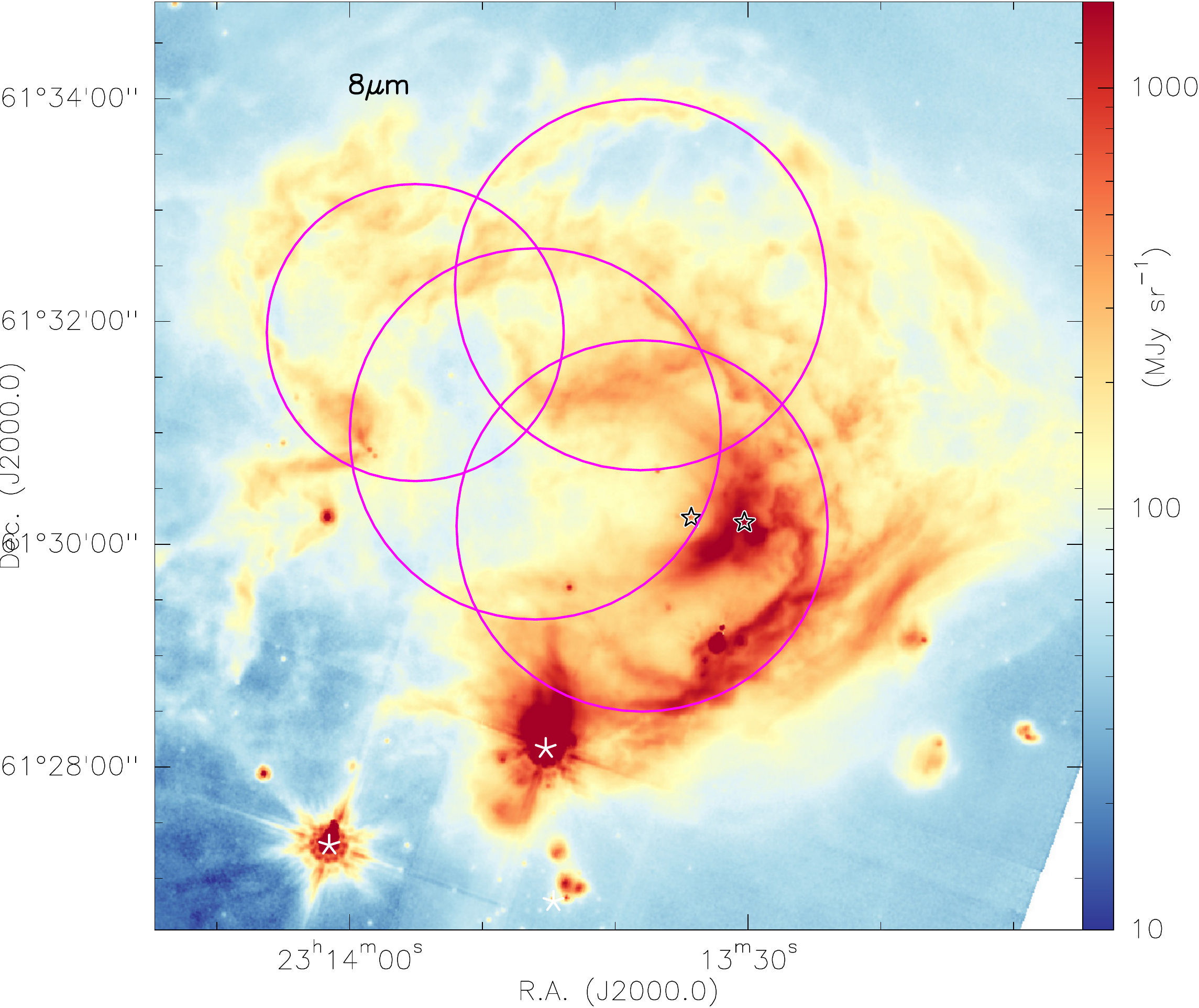}
\caption{The color-scale shows the 8\,$\mu$m emission, and the purple circles again outline the tentative bubble structures.}
\label{8mu_bubbles} 
\end{figure}

For comparison, the ring/bubble candidates are also plotted on the
8\,$\mu$m emission that mainly stems from PAHs emitted in PDRs
(Fig.~\ref{8mu_bubbles}). Several of the ring/bubble-like structures
can clearly be seen in this PAH emission.

The suggested bubble A is best visible as limb-brightened rings in the
channels between -65 and -59\,km\,s$^{-1}$. An additional indicator
that this bubble may indeed be a real physical structure is that at
the most blue-shifted velocities (channel at -69\,km\,s$^{-1}$), the
emission is mainly at the center of this proposed bubble A. This is
exactly what one expects when this bubble is expanding along the line
of sight (see also appendix \ref{bubble_finding}), and was shown also
for RCW120 \citep{luisi2021}.

\begin{figure*}[htb]
\includegraphics[width=0.99\textwidth]{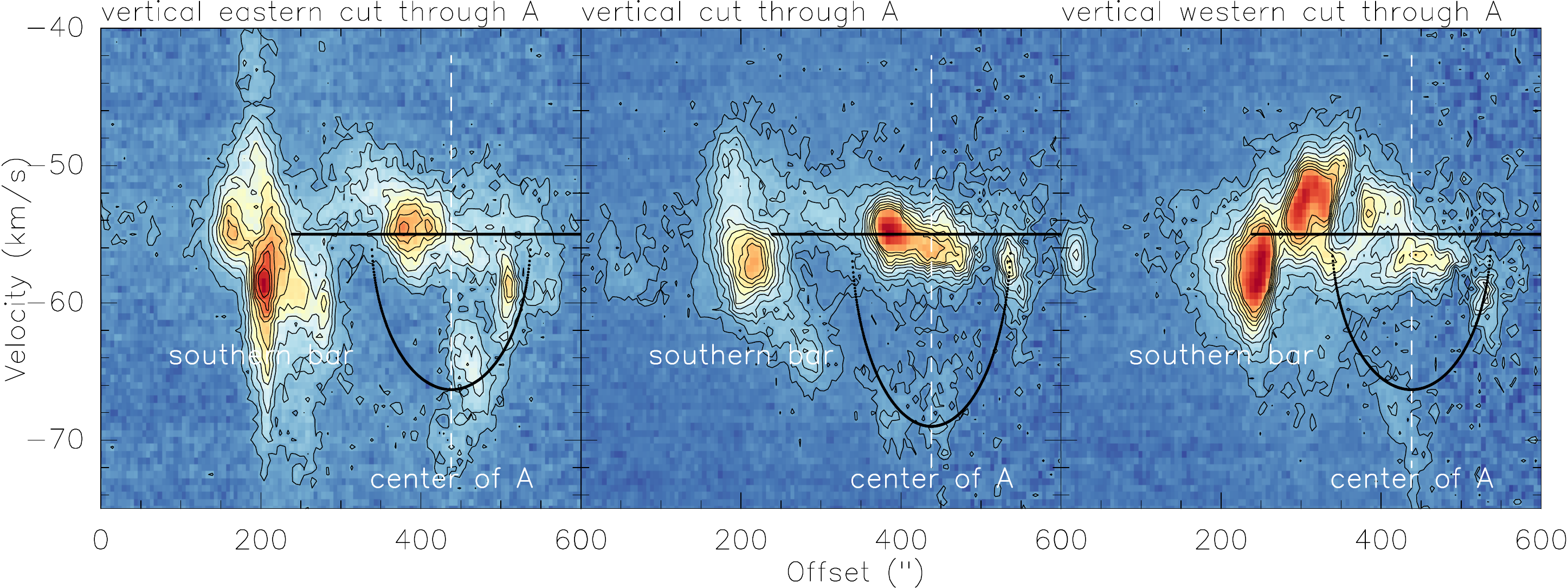}
\caption{Vertical position-velocity cuts through bubble A from south
  to north derived from the [CII] line emission. The central panel
  goes through the estimated central position of A whereas the left
  and right panels are shifted $1'$ to the east and west,
  respectively. The white dashed line marks the approximate center of
  the bubble. The black dotted lines in the middle panel outline the
  elliptical shape a spherical shell with radius $100''$ and expansion
  velocity of 14\,km\,s$^{-1}$ would have in the pv diagram. The
  dotted lines in the left and right panel correspond to the same
  expanding shell at the $1'$ shift to the east and west where the
  $1'$ corresponds to $\sim$36\,deg offset in the sphere (e.g.,
  \citealt{butterfield2018}, sketch in their appendix A.2). Hence the
  velocities are then reduced by $cos(36^{\circ})$.}
\label{pv_bubble_A} 
\end{figure*} 

\paragraph{Position-velocity analysis:} A different way to visualize the velocity structure of bubble A is via
a position-velocity cut. Figure \ref{pv_bubble_A} presents three
vertical pv-diagrams in the [CII] line through the center of
bubble A and shifted $1'$ to the east and west,
respectively. Corresponding horizontal pv-diagrams are presented in
the appendix in Fig.~\ref{pv_bubble_A_hor}.  While most velocities
along these cuts are between $-50$ and $-60$\,km\,s$^{-1}$, there is
clearly high-velocity blue-shifted gas around $\sim -70$\,km\,s$^{-1}$
close to the center of the bubble, indicating an expanding structure
along the line of sight. Figures \ref{pv_bubble_A} and
\ref{pv_bubble_A_hor} also show the shape a spherical shell with a
radius if $100''$ and an expansion velocity of 14\,km\,s$^{-1}$ would
have in such a pv-diagram. This is not a fit but just outlining how
such a shell would appear as an ellipse in the pv-diagram. The left
and right panels with pv-cuts at offsets of $1'$ correspond in a shell
geometry to angles of $\sim$36\,deg (see a sketch of such a geometry
in appendix A.2 from \citealt{butterfield2018}). The velocities at
such angle are then reduced by $cos(36^{\circ})$. While the data and
toy model at negative blue-shifted velocities show some resemblance of
each other, consistent with a spherical half-shell, the
red-shifted side at positive relative velocities shows no indication
of such a shell-like geometry. This could either indicate that such a
shell model is insufficient, or the center of the shell could be close
to the rear side of the cloud that only blue-shifted gas moving
towards the observer can be detected while the red-shifted expansion
leaves the cloud almost immediately that no gas gets accelerated to a
detectable level. In that context, \citet{luisi2016} estimated that
approximately 15\% of the ionized gas is leaking outside of the H{\sc
  ii} region.

\begin{figure*}[htb]
\includegraphics[width=0.99\textwidth]{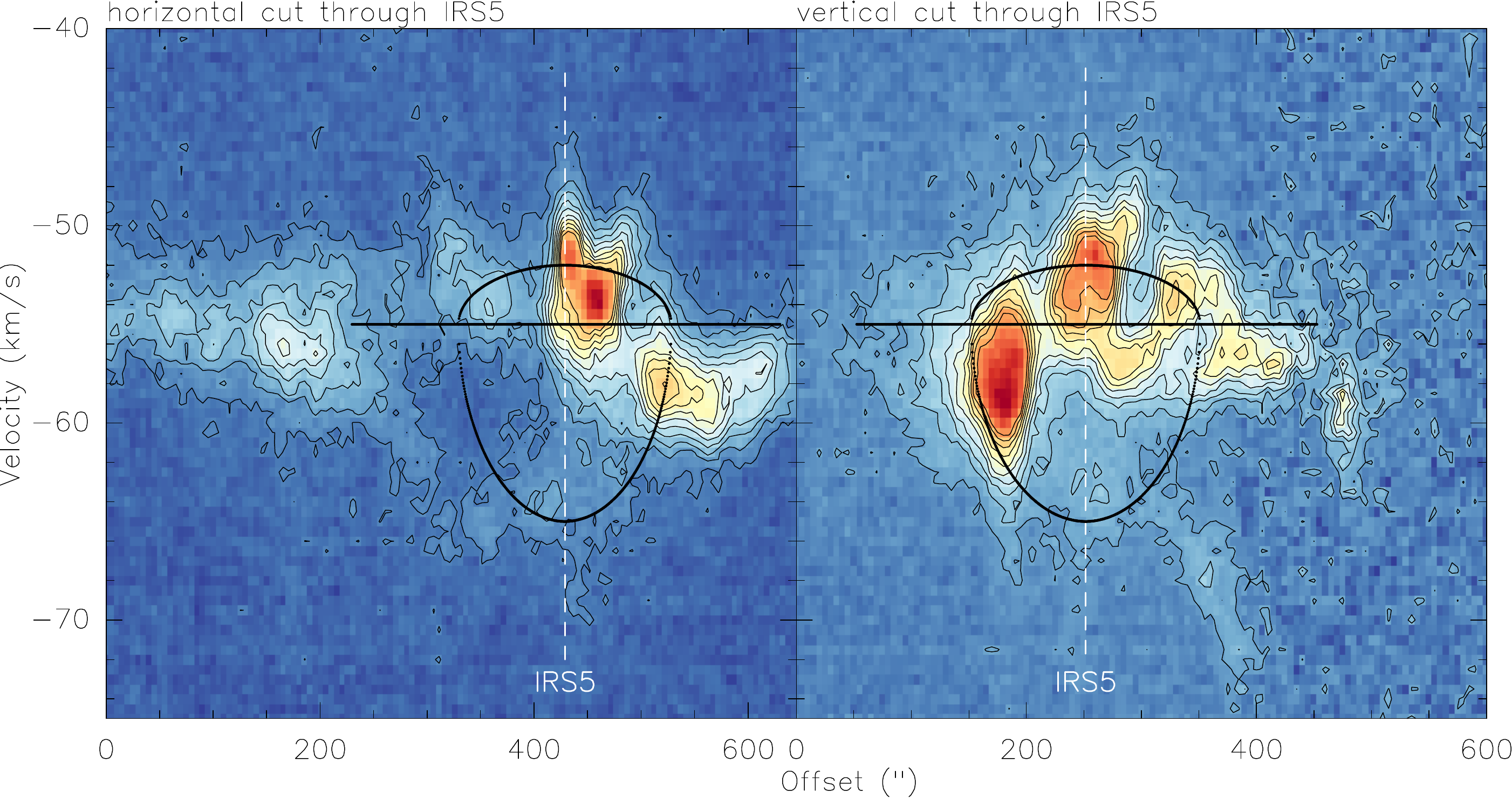}
\caption{Horizontal (along R.A., left) and vertical (along Dec.,
  right) position-velocity diagrams through the position of one of the
  exciting sources IRS5 (marked in each panel) derived from the
    [CII] line emission. corresponding roughly to the estimated
  center of bubble B. The contour levels are in $3\sigma$ steps. The
  black dotted lines outline the elliptical shape a spherical shell
  with radius $100''$ and expansion velocities of 10\,km\,s$^{-1}$ and
  3\,km\,s$^{-1}$ at negative and positive relative velocities,
  respectively, would have in the pv diagram.}
\label{pv_bubble_B} 
\end{figure*}

The structure B is centered close to the main excitation sources in
the region, IRS5 and IRS6 (Fig.~\ref{dss2}). But we note that the
automated bubble-identification approach does not center it exactly on
the excitation sources but a bit shifted. Bubble B's border at the
south-western side is roughly the bar-like PDR discussed in section
\ref{pdr}. In the north, structures visible in particular at
velocities of $-55$/$-53$\,km\,s$^{-1}$ delineate the potential sphere
of that bubble. Doing position-velocity cuts in the horizontal as well
as vertical direction through this bubble B and source IRS5
(Fig.~\ref{pv_bubble_B}), one also finds that the highest-velocity gas
on the blue- as well as red-shifted side is found close to IRS5. This
is again indicative of expanding motions along the line of side,
potentially caused by IRS5 and IRS6. We again draw the corresponding
shapes for expanding shells into the pv-diagram
(Fig.~\ref{pv_bubble_B}). While the radius of 100$''$ is given by our
shell-identification approach above, the velocities on the blue- and
red-shifted side differ significantly. To outline those differences,
we used an expansion velocity of 10\,km\,s$^{-1}$ for the blue-shifted
side, and 3\,km\,s$^{-1}$ for the red-shifted side. If the observed
structure really belongs to an expanding shell, such velocity
differences would indicate a highly inhomogeneous medium where the
red-shifted expansion would enter into much denser material. Another
option is that the central velocity is not $-55$\,km\,s$^{-1}$ but a
few km\,s$^{-1}$ higher. In that case, we would just have a
blue-shifted bubble-structure without any red-shifted counterpart,
similar to structure A discussed before. That would put the center of
structure B also on the rear side of the cloud.

\begin{figure*}[htb]
\includegraphics[width=0.99\textwidth]{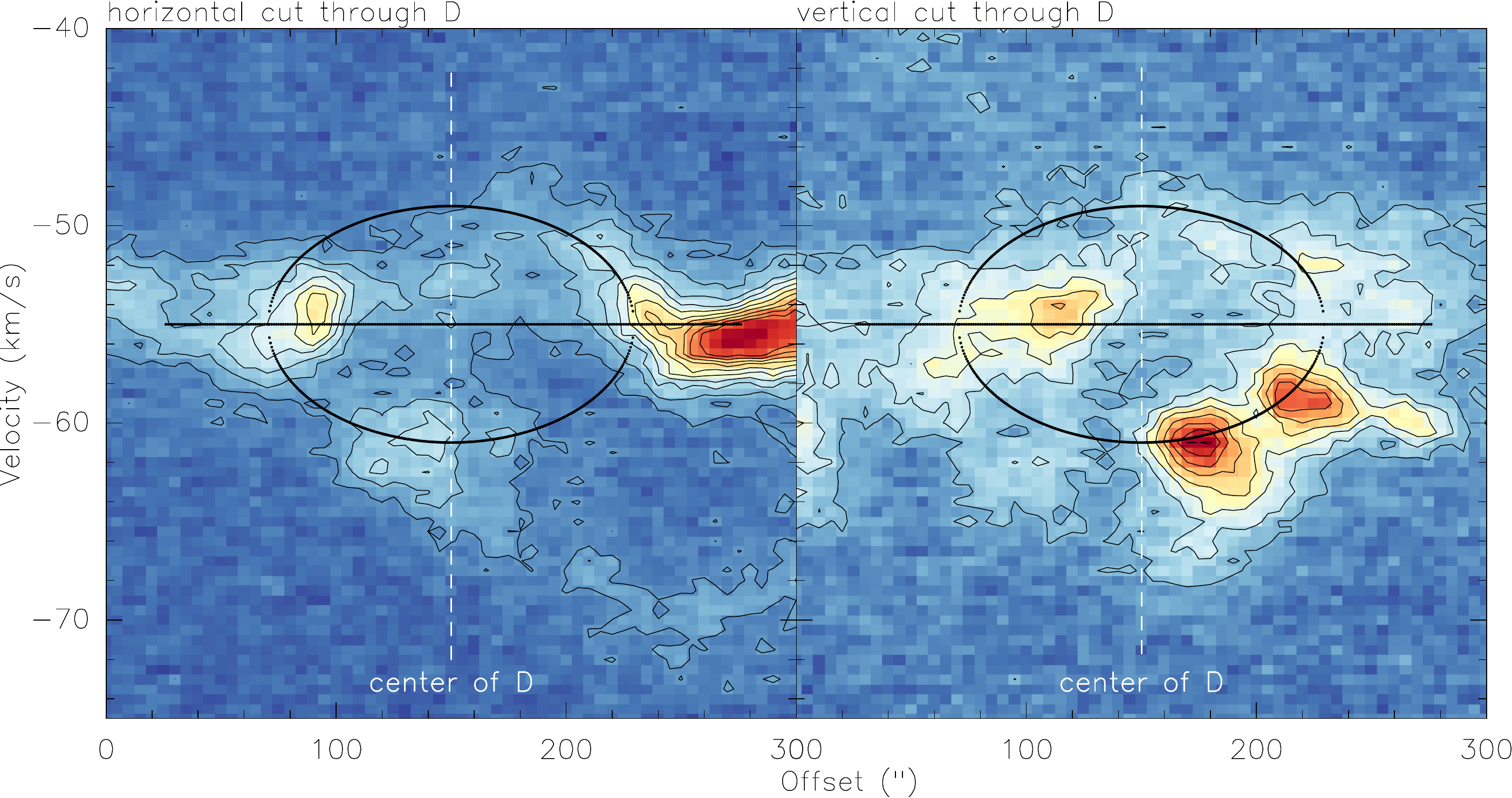}
\caption{Horizontal (left) and vertical (right) position-velocity
  diagrams through the center of bubble D (marked in each panel)
  derived from the [CII] line emission. The contour levels are in
  $3\sigma$ steps. The black dotted lines outline the elliptical shape
  a spherical shell with radius $80''$ and expansion velocities of
  6\,km\,s$^{-1}$ would have in the pv diagram.}
\label{pv_bubble_D} 
\end{figure*} 

The identified structure C is in-between the other bubble/ring
structures. While it may be a real physical entity, it could also
arise from an overlap of the other bubble/ring structures. Therefore,
we omit further detailed discussion for that structure.

Furthermore, we identify structure D where the limb-brightened
ring-like features are best delineated in the channels between $-57$ and
$-53$\,km\,s$^{-1}$. In addition to the bubble edges, structure D again
shows high-velocity gas at its center, typically signposting expanding
structures. These centered emission structures in bubble D are best
visible in the blue-shifted channels at $-67$/$-65$\,km\,s$^{-1}$ and in
the most red-shifted channels at $-49$/$-47$\,km\,s$^{-1}$. For
comparison, Fig.~\ref{pv_bubble_D} presents the corresponding vertical
and horizontal pv-cuts through the central position of structure D. In
contrast to the pv-diagrams of bubbles A and B
(Figs.~\ref{pv_bubble_A} and \ref{pv_bubble_B}) that show the
high-velocity gas mainly at negative blue-shifted velocities, the
pv-diagram of structure D, in particular the one in vertical direction
(right panel in Fig.~\ref{pv_bubble_D}), shows an almost ring-like
structure with high velocities on the blue- and red-shifted side, in
good agreement with the channel-map discussed before. For comparison,
we again show the shape an expanding shell with radius $80''$ and an
expansion velocity of 6\,km\,s$^{-1}$.

\paragraph{Driving sources:} It is interesting to note that the ring-like structure of high
velocity dispersion in the [CII] moment map (Fig.~\ref{moments})
roughly coincides with the combined outline of the four bubble/ring
structures discussed here (Fig.~\ref{channel_cii}). Since the high
velocity dispersion in Fig.~\ref{moments} largely stems from several
velocity components (Fig.~\ref{spectra}), these velocity signatures
may stem from expanding motions from the H{\sc ii} region. While for
bubble B, the driving source for bubble expansion may be the main
exciting sources of the H{\sc ii} region (IRS5 and IRS6,
\citealt{puga2010}), this is less obvious for the other structures A,
C and D. For these three structures, we have not identified a clear
central source that could be associated with bubble expansion. While
\citet{puga2010} identified more young stellar objects in the
environment of the H{\sc ii} region, they are not at the bubble
centers (which in fact IRS5 and IRS6 are not exactly either), and
furthermore are typically of lower luminosity. Also Spitzer data do
not allow us to identify clear driving sources.

While the O3 and O9 stars IRS6 and IRS5 \citep{puga2010} are most
likely the main exciting sources of the H{\sc ii} region, the
identification of several more ring- or bubble-like structures does
not necessarily mean that there are more bubble-driving sources in the
field. While proper motion of the exciting stars may cause the
dislocation of IRS5 and IRS6 from the center of bubble-structure B,
such displacement may also be caused by density and pressure gradients
in the region. A similar displacement is also observed between the
Trapezium stars and the main bubble in Orion, the Orion Veil
\citep{,pabst2019,pabst2020}.

\paragraph{Multiple bubbles versus expansion into an inhomogeneous medium:} In addition to this, the strong sub-structure of the H{\sc ii} region
visible in the different data sets, in particular in the [CII] channel
map, indicates an almost porous structure of the H{\sc ii} region
where radiation and wind energy can leak out and influence also parts
of the H{\sc ii} region that are less close to the main exciting
sources IRS5 and IRS6. Hence, while bubble-like structures are
possible, the NGC7538 H{\sc ii} region may also expand into a very
inhomogeneous environment. If an H{\sc ii} region expands in such an
inhomogeneous cloud, changes of the cloud morphology are
expected. Such impact may also cause bubble-like morphologies as
observed here. Recent analytic and numerical work by
\citet{lancaster2021a,lancaster2021b,lancaster2021c} also showed that
the wind-driven expansion of the gas around H{\sc ii} regions can
cause different morphological sub-structures. For the NGC7538 H{\sc
  ii} region, a clear discrimination between several bubble-structures
each driven by individual internal sources and/or an expansion of
mainly one wind-driven shell into an inhomogeneous medium cannot be
drawn. Further hydrodynamic modeling of expanding H{\sc ii} regions
into an inhomogeneous medium may show whether such multiple ring-like
structures of expanding gas can occur or whether separate
bubble-driving centers are required.

\paragraph{Thermal expansion or winds of the high-mass stars:}
Independent of whether all the ring-like structures are real bubbles
or whether they are rather caused by an expansion of the gas into an
inhomogeneous medium, the high-velocity components visible in the
pv-diagrams (Figs.~\ref{pv_bubble_A} to \ref{pv_bubble_D}) all show
high-velocity gas likely driven by the expansion of the H{\sc ii}
region. In particular the gas towards structures A and B exhibit
expansion velocities $\geq 10$\,km\,s$^{-1}$. Taking an expansion
velocity of 10\,km\,s$^{-1}$ and a radius of $100''$ at face-value,
this would correspond to an expansion time-scale of roughly
0.126\,Myrs. This is comparably short with respect to the
observationally estimated ages of the cluster. For example,
\citet{puga2010} derived an age range between 0.5 and 2.2\,Myrs, and
\citet{sharma2017} estimated a mean age of the young stellar objects
of 1.4\,Myrs with a range between 0.1 and 2.5\,Myrs. Hence, a constant
expansion at that velocity is unlikely the real scenario for that
region. However, can the measured expansion velocities be caused by
thermal expansion of the gas or by the winds of the high-mass stars?

To estimate expected expansion velocities, for the thermal expansion
we follow the Spitzer solution \citep{spitzer1998}, whereas for the
wind-driven expansion, the approach by \citet{weaver1977} and its
adaption to include radiative cooling by \citet{maclow1988} is
adopted. A comprehensive summary of these two approaches is given in
\citet{henshaw2021}. The expansion velocities depending on time for
the thermal expansion $v_{th}(t)$ and the wind-driven expansion
$v_{w}(t)$ are:

$$v_{th}(t) = c_s \left(1+\frac{7c_s t}{46} \right)^{-3/7}$$

$$v_w(t) = \frac{1}{2}\frac{125}{154\pi}\left( \frac{L_w}{\rho}\right)^{1/5}t^{-1/2} t_{cool}^{1/10}.$$

Here, $c_s$ is the sound speed in the ionized gas (8\,km\,s$^{-1}$ at
5000\,K), $L_w=\frac{1}{2}\dot{M}v_{\infty}^2$ is the mechanical wind
luminosity, $\rho$ the ambient density, and $t_{cool}$ the cooling
time following \citet{maclow1988} and \citet{henshaw2021}. The wind
mass flow rate $\dot{M}$ and the escape velocity $v_{\infty}$ for an
O3 star (IRS6, \citealt{puga2010}) are taken from
\citet{muijres2012}. The wind-driven velocities $v_w$ strongly depend
on the density of the ambient gas $\rho$ (also $t_{cool}$ depends on
$\rho$, \citealt{maclow1988}).

Following \citet{puga2010} and \citet{sharma2017}, the age of the
NGC7538 cluster and associated young stellar objects is between
roughly 0.1 and 2.5\,Myrs. Even at the lower boundary of 0.1\,Myr, the
highest expansion velocities one can get with a Spitzer-type thermal
expansion is $\sim$5.7\,km\,s$^{-1}$, far below what is measured in
NGC7538. So, purely thermal expansion cannot properly explain the
measured high-velocity gas. In contrast to that, the wind solution
gives significant higher velocities. For the given parameters, and
assuming densities of $10^3$, $2\times 10^3$ and $3\times
10^3$\,cm$^{-3}$, expansion velocities around 10\,km\,s$^{-1}$ are
estimated at times of roughly 1.1, 0.36 and 0.17\,Myr,
respectively. While the lower end would be more consistent with the
constant expansion estimated above, the higher time-scales are more
consistent with the the estimated age limits by \citet{puga2010} and
the mean ages of the young stellar objects derived by
\citet{sharma2017}. Hence, wind-driving seems a plausible way to
explain the observed high-velocity gas. The wind-driving is also
consistent with the diffuse X-ray emission towards that region that is
typically attributed to hot plasma caused by the wind shocks of
the massive stars (e.g., \citealt{guedel2008,townsley2018,pabst2020}).

Regarding the potential discrepancy between the lower end of the
estimated cluster age at 0.5\,Myrs and the shorter time-scales
estimated for either constant expansion or the wind expansion at
higher densities, one way to reconcile these is that initially the
bubble may have confined inside the dense molecular core in which IRS5
and IRS6 have formed. Only once the H{\sc ii} region broke out of this
core, expansion was rapid into the surrounding lower density
material. In this picture, the expansion timescale one estimates then
refers to the time since the breakout from this dense core. 

\subsection{The bar-like photon dominated region (PDR)}
\label{pdr}

\begin{figure}[htb]
\includegraphics[width=0.49\textwidth]{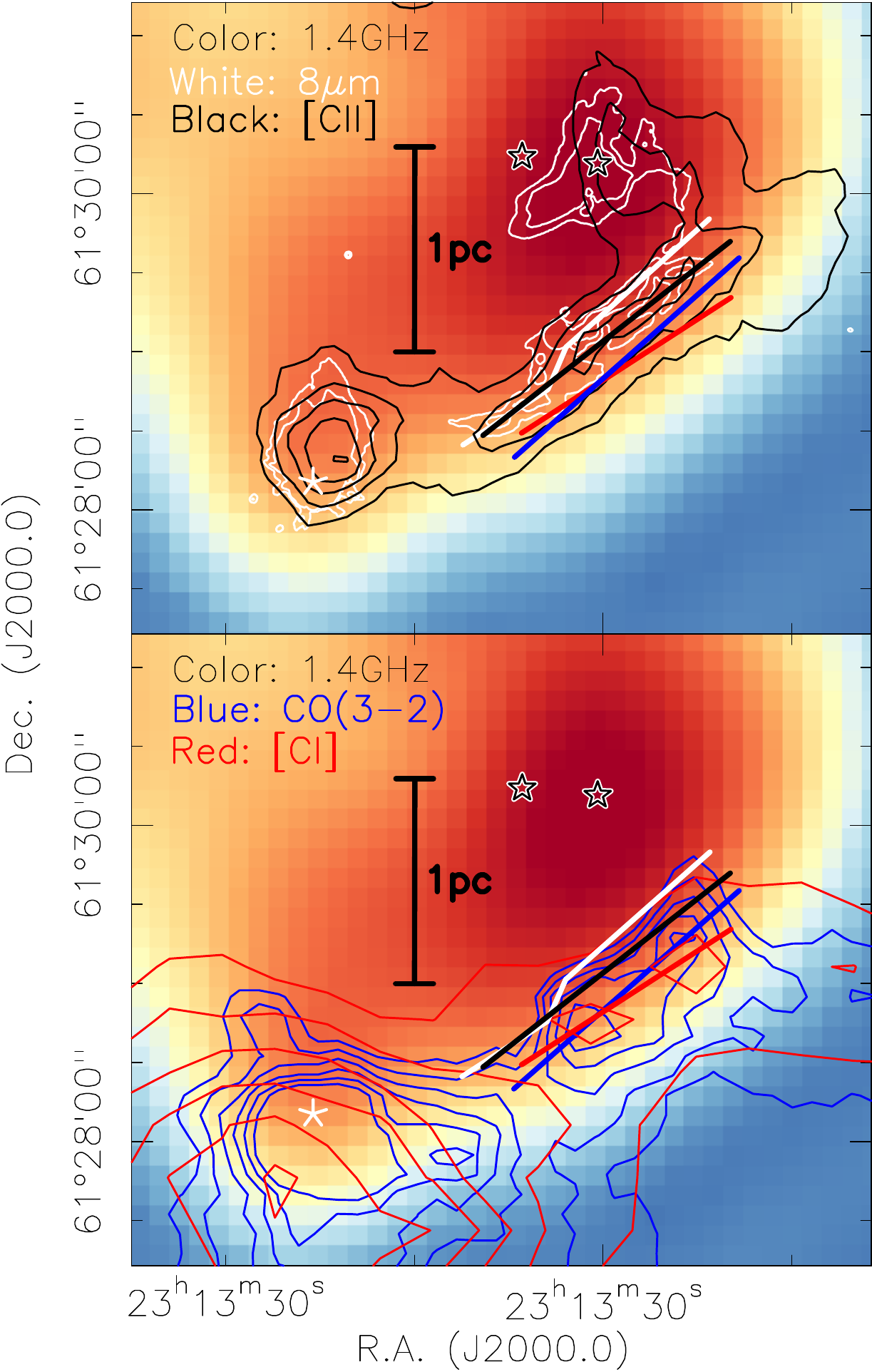}
\caption{Zoom on the bar-region south of the exciting sources IRS5 and
  IRS6. The color-scale and the colored contours show the emission
  tracers labeled in the two panels. Only selected contours are shown for
  clarity. The four colored lines mark the approximate positions of
  the emission crests in the corresponding tracers. The sources IRS5,
  IRS6 and IRS1 are marked. A linear scale-bar is shown as well.}
\label{bar} 
\end{figure}

Figure \ref{bar} presents a zoom into the bar-like structure at the
south-western edge of the expanding H{\sc ii} region. While the
color-scale shows the ionized gas as traced by the 1.4\,GHz continuum
emission, the contours outline various other tracers: 8\,$\mu$m
continuum, [CII], CO(8--7) \& (3--2) and [CI] emission. The white,
black, blue and red lines mark the approximate locations of the
emission crests in the 8\,$\mu$m, [CII], CO(3--2) and [CI] emission,
respectively. The crests were identified by eye via connecting
  the emission peaks. The crest of the CO(8--7) emission
approximately coincides with that of the [CII] emission.

In this picture, the 8\,$\mu$m emission, that should stem largely from
UV-pumped infrared fluorescence PAHs, peaks closest to the exciting
sources of the H{\sc ii} region. Second in this layered structures is
the ionized carbon [CII] emission that spatially approximately
coincides with the highly excited CO(8--7) emission. Furthest away
from the exciting sources are the emission crests of the lower excited
CO(3--2) and [CI] emission that, given the spatial resolution of the
data, projected on the plane of the sky spatially approximately
coincide. This is different around the young high-mass star-forming
region IRS1 south-east of the bar where CO(3--2) peaks closer to IRS1
than the atomic carbon [CI] line (Fig.~\ref{bar}). Coming back to the
bar-like structure, in this framework, we can separate the photon
dominated region (PDR) in roughly four layers:\\

\noindent (i) The exciting sources and 1.4\,GHz continuum emission\\
(ii) the 8\,$\mu$m emission, likely PAHs\\
(iii) ionized carbon [CII] and highly excited CO(8--7)\\
(iv) atomic carbon [CI] and lower excited CO(3--2)\\

This layered structure resembles prototypical edge-on PDR structures
observed also for example in the Orion bar (e.g.,
\citealt{tielens1993,tauber1994,tauber1995,simon1997b}) or M17SW
(e.g., \citealt{perez2015,perez2015b}). We will return to the carbon
budget within this PDR in Section \ref{carbon_budget}.

\begin{figure}[htb]
\includegraphics[width=0.49\textwidth]{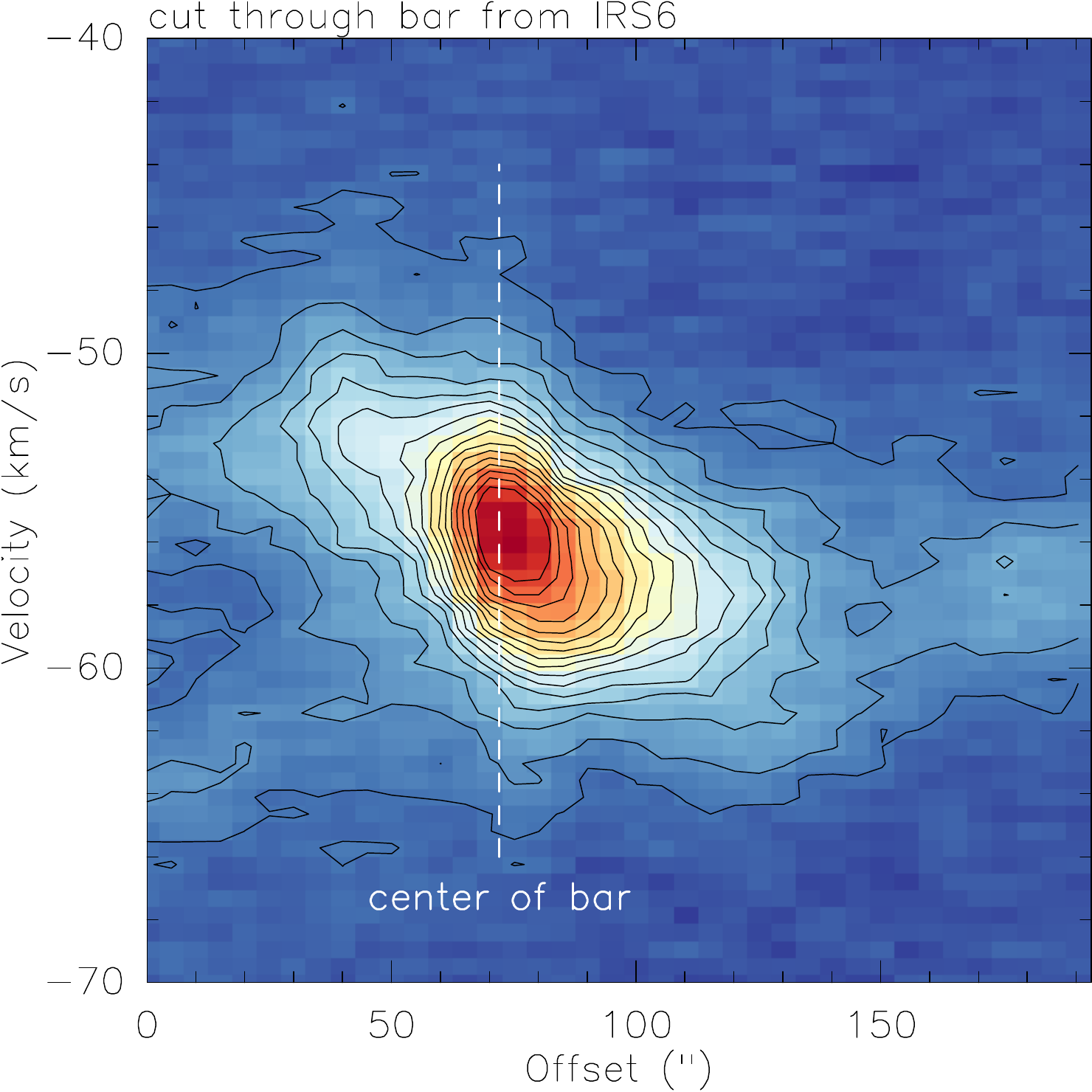}
\caption{Position-velocity cut from the [CII] line from IRS6 through
  the bar as outlined in the middle panel of Fig.~\ref{moments}. The
  white dashed line marks the approximate center of the bar.}
\label{pv_bar} 
\end{figure} 

In addition to the layered structure, the 1st moment map in
Fig.~\ref{moments} indicates that there may be a velocity gradient
across the bar. To get a better view of that, Fig.~\ref{pv_bar}
presents a position-velocity diagram across the bar starting from the
O3 star IRS6. We see a clear velocity gradient from $\sim
-50$\,km\,s$^{-1}$ at the beginning of the cut near the main exciting
source IRS6 to blue-shifted velocities $< -60$\,km\,s$^{-1}$ on the
other side of the bar. The fact that the gas within the bar gets
blue-shifted indicates that the bar lies along the line of sight to
the observer in front of the exciting source IRS6.

\subsection{Carbon budget}
\label{carbon_budget}
 
\begin{figure}[htb]
\includegraphics[width=0.49\textwidth]{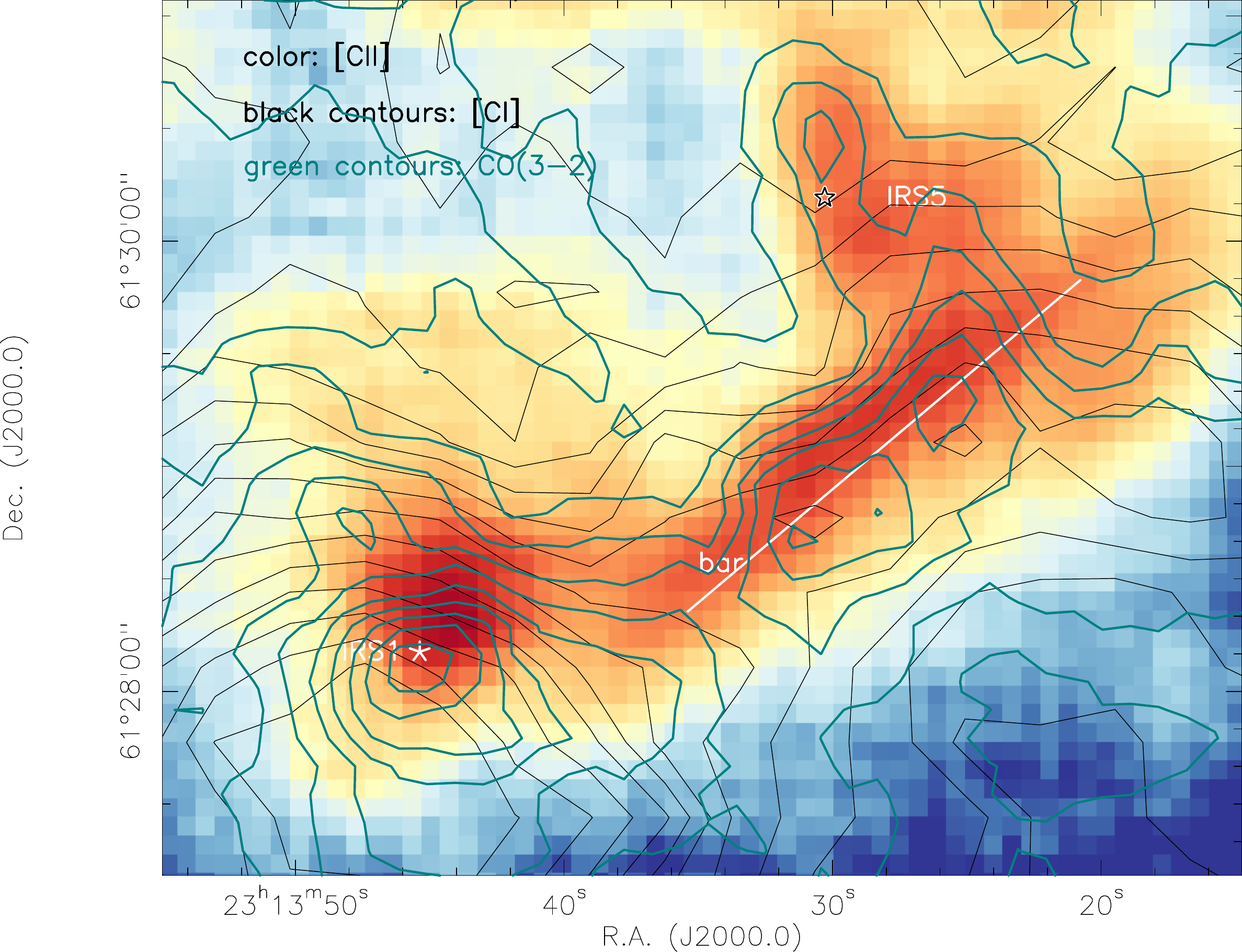}
\caption{Integrated [CII] (color), [CI] (black contours) and CO(3--2)
  (green contours) emission in the bar-region where the [CII]/[CI]/CO
  mass estimates are conducted (section \ref{carbon_budget}). All
  three maps were integrated between -70 and -40\,km\,s$^{-1}$. The
  [CI] contours start at $6\sigma$ and continue in $3\sigma$ steps
  ($1\sigma=1.6$\,K\,km\,s$^{-1}$). The CO(3--2) contours start at
  $10\sigma$ and continue in $10\sigma$ steps
  ($1\sigma=3.2$\,K\,km\,s$^{-1}$). The white line marks the
  approximate location of the bar-like feature.}
\label{carbon} 
\end{figure}

While carbon is mainly in its ionized C$^+$ state in large fractions
of the PDR, it is mainly in its molecular CO form in the dense
molecular cloud (Fig.~\ref{mom0}). But what is the relative carbon
budget within the transition zone from the H{\sc ii} region towards
the molecular clouds? This transition can best be studied in the
south-western bar-like structure. Figure \ref{carbon} presents the
ionized carbon [CII], the atomic carbon [CI] and the molecular carbon
CO(3--2) emission toward that region. The integration regime for all
three species is the same from $-70$ to $-40$\,km\,s$^{-1}$. In the
following, we derive approximate estimates for the column densities
and masses of the different carbon components in the bar region
outlined by Fig.~\ref{carbon}. With the assumptions outlined below,
these should only be considered as rough estimates. But nevertheless,
they allow us to asses roughly how much mass is within each carbon
component in and around the bar-like region.

The column densities of the ionized carbon are estimated from the
[CII] line following \citet{goldsmith2012}, eq.~26, adding an
additional optical depth term of $ \frac{\tau}{1-e^{-\tau}}$ to
correct for the optical depth $\tau$ \citep{rohlfs2006}:

$$  N_{\rm{[CII]}} = \frac{1}{3.43\times 10^{-16}}\left[1+0.5e^{91.25/T_{\rm{kin}}}\left(1+\frac{2.4\times 10^{-6}}{C_{ul}}\right)\right]$$
$$ \times  \frac{\tau}{1-e^{-\tau}} \int T_{\rm{mb}}dv ~\rm{[cm^{-2}]}$$

with the kinetic temperature $T_{\rm kin }$ (see further below), the
collision rate $C_{ul}$, the intensity $T_{\rm{mb}}$ in K and the
velocity $dv$ in km\,s$^{-1}$. The collision rate $C_{ul}=R_{ul}n$
depends on the temperature, where $R_{ul}$ is collision rate
coefficient with H$_2$ and $n$ the density, assumed to be around
$\sim$10$^3$\,cm$^{-3}$.  The collision rate coefficients $R_{ul}$
with H$_2$ are taken from the Leiden database for molecular
spectroscopy (\citealt{schoeier2005},
http://home.strw.leidenuniv.nl/~moldata/, \citealt{lique2013}) where
the para- and ortho-rates are weighted following \citet{lebourlot1991}
and \citet{gerlich1990}. The comparably good spatial correspondence of
the [CII] with the molecular line emission around the bar-shaped PDR
indicates that H$_2$ should be the most dominant collisional
partner. If for comparison all collisions were with atomic hydrogen,
with the given assumptions we would get masses roughly a factor of
1.57 lower (see also \citealt{goldsmith2012}). In reality, it is
likely a mix of collisional partners dominated by H$_2$ and hence the
difference should be much smaller. In the following we use the
collisional rates with H$_2$. Evaluating the [$^{13}$CII] emission in
our data, we detect the line after averaging over areas to decrease
the noise. In the case of the bar-like region in NGC7538, the
[$^{13}$CII] data indicate a mean optical depth $\tau$ for the main
[CII] line of $\sim$2.5. Compared to the optically thin case, this
corresponds to a column density and mass correction factor of
$\sim$2.7.

Furthermore, the atomic carbon column densities are estimated from the
[CI] emission following \citet{frerking1989}:

$$ N_{\rm{[CI]}} = 5.94\times 10^{15} \frac{1+3e^{\frac{-23.6}{T_{\rm{ex}}}}+5e^{\frac{-62.4}{T_{\rm{ex}}}}}{3e^{\frac{-23.6}{T_{\rm{ex}}}}} \int T_{\rm{mb}}dv ~\rm{[cm^{-2}]}$$

with the excitation temperature $T_{\rm{ex}}$ (the temperature
quantification is discussed further below).  The CO column density can
then approximately be estimated following standard equations (e.g.,
\citealt{cabrit1988}):
  
$$N_{\rm{CO}} = \frac{6.97\times 10^{15}}{\nu^2 \mu^2} T_{\rm{ex}}e^{E_{\rm{u}}/(kT_{\rm{ex}})} \frac{\tau}{1-e^{-\tau}}$$
$$ \times \int T_{\rm{mb}}(^{12}\rm{CO(3-2)})dv ~\rm{[cm^{-2}]}.$$

Here, $\nu$, $\mu$, $T_{\rm{ex}}$, $E_{\rm{u}}/k$, $\tau$ and
$T_{\rm{mb}}$ are the frequency, dipole moment (0.112\,Debye), the
excitation temperature, the upper energy level of 33.19\,K, the
optical depth and the main beam brightness temperature of the line,
respectively.

Following \citet{cabrit1988}, assuming $^{13}$CO is optically thin, the
optical depth term can be approximated as:

$$\frac{\tau}{1-e^{-\tau}} = \frac{[^{12}\rm{CO}]}{[^{13}\rm{CO}]}\left(\frac{\int T_{\rm{mb}}(^{13}\rm{CO})dv}{\int T_{\rm{mb}}(^{12}\rm{CO})dv}\right)$$

The term $\frac{[^{12}\rm{CO}]}{[^{13}\rm{CO}]}$ corresponds to the
$^{12}$CO/$^{13}$CO abundance ratio. According to
\citet{moscadelli2009}, NGC7538 is in the Perseus arm which results at
the given longitude of the region (111.5\,deg) in a Galactocentric
distance of $\sim$9\,kpc (e.g., \citealt{reid2019}). Following
\citet{wilson1994}, the $^{12}$CO/$^{13}$CO abundance ratio is then
$\sim$75.  Furthermore, the term $\left(\frac{\int
  T_{\rm{mb}}(^{13}\rm{CO})dv}{\int
  T_{\rm{mb}}(^{12}\rm{CO})dv}\right)$ corresponds to the mean ratio
of brightness intensities between the $^{13}$CO and the $^{12}$CO
transition. Within the IRAM NOEMA large program CORE
\citep{beuther2018b}, a small area of $1.5'\times 1.5'$ around IRS1
had been mapped in the $J=2-1$ transitions of $^{12}$CO and $^{13}$CO
with the IRAM 30\,m telescope. Using these small-area data, we
estimate a mean ratio between those two lines of $\sim 0.2$ in the
velocity regime between $-70$ and $-40$\,km\,s$^{-1}$. This value is
used in the following also for the here observed $J=3-2$
transition. This results in the column density equation:

$$N_{\rm{CO}} = \frac{6.97\times 10^{15}}{\nu^2 \mu^2} T_{\rm{ex}}e^{E_{\rm{u}}/(kT_{\rm{ex}})} \frac{[^{12}\rm{CO}]}{[^{13}\rm{CO}]}\left(\frac{\int T_{\rm{mb}}(^{13}\rm{CO})dv}{\int T_{\rm{mb}}(^{12}\rm{CO})dv}\right)$$
$$ \times \int T_{\rm{mb}}(^{12}\rm{CO(3-2)})dv ~\rm{[cm^{-2}]}.$$

In the following, we will calculate the mean column densities and the
corresponding masses over the area presented in Figure \ref{carbon}
that encompassed roughly 11.25\,pc$^{2}$.  The main missing ingredient
in the above equations are the temperatures to be chosen for the column
density estimates.

Regarding the ionized carbon C$^+$, \citet{langer2010} typically
assume temperatures between 100 and 150\,K, which seems a reasonable
temperature regime in this PDR as well. Using these two values as
brackets for the possible temperatures of the ionized carbon C$^+$ we
find an ionized carbon C$^+$ mass of $M_{\rm{[CII]}}\approx
9.7\pm 1.6$\,M$_{\odot}$, assuming the above discussed mean optical
depth $\tau\sim 2.5$. This is an upper limit to the mass
\citep{rohlfs2006} whereas we can consider an estimate assuming
optically thin emission as a lower limit. In the optically thin limit,
we get the lower limit of $M_{\rm{[CII]\_lower\_limit}}\approx 3.6\pm
0.6$\,M$_{\odot}$, a factor $\sim$2.7 below the optical depth
corrected mass estimate.

For the atomic and molecular gas, the excitation temperatures are not
well determined in that region. However, variations in the excitation
temperatures between 30 and 50\,K do not vary the estimated masses
significantly. Therefore, we estimate the atomic $M_{\rm{[CI]}}$ and
molecular masses $M_{\rm{CO}}$ in that excitation temperature
regime. In that framework we find $M_{\rm{[CI]}}\approx 0.45\pm
0.1$\,M$_{\odot}$ and $M_{\rm{CO}}\approx 1.2\pm
0.1$\,M$_{\odot}$. The molecular mass can also be compared to the gas
mass one can derive from the Herschel gas column density map. While a
H$_2$-to-CO ratio of $10^4$ would result in a gas mass of $\sim
1.2\times 10^4$\,M$_{\odot}$, the gas mass derived from the Herschel
map in the same area is $\sim$4600\,M$_{\odot}$. This difference
around a factor 2.6 can already be explained by the uncertainties of
temperatures and line-of-sight averaging, optical depth estimates and
the assumed dust properties for the Herschel-based mass
estimate. Nevertheless, this comparison shows that our molecular mass
estimate has an approximate uncertainty of a factor two.

While atomic C$^0$ and molecular CO carbon have comparably low masses
between 0.45 and 1.2\,M$_{\odot}$, the mass of the ionized carbon
C$^+$ is in the optically thin lower limit a factor $\sim$3--8 larger,
and for the optical depth corrected mass estimate even a factor
$\sim$8--22 larger in this bar-like PDR (Fig.~\ref{carbon}). Hence,
the ionized carbon C$^+$ is clearly the dominant carbon form in this
PDR. As expected, this is very different to the previous carbon
studies in infrared dark clouds where the carbon was found to be
dominantly in the molecular CO form \citep{beuther2014}.

\subsection{Decoupled [CII] component}
\label{decoupled}

The [CII] component at velocities below $-60$\,km\,s$^{-1}$ deserves
separate discussion because it is identified mainly in the [CII]
emission without strong counterparts in the molecular and atomic gas
(Figs.\,\ref{spectra} \& \ref{channel_cii}). A spectral feature
dominantly visible in [CII] emission has rarely been identified in the
past in regions of expanding H{\sc ii} regions and feedback. For
comparison, the veil in Orion exhibits barely any extended CO
emission, however, compact globules are identified within the [CII]
emitting gas structures \citep{goicoechea2020}.

Inspecting the channel map with proposed bubble-like structures
(Fig.~\ref{channel_cii}) in velocity channels at $\leq
-65$\,km\,s$^{-1}$, this blue-shifted high-velocity gas appears to be
associated with all four identified bubble-structures as well as with
the young star-forming region IRS1 in the south of the region. Since
IRS1 is known to be an active driver of molecular outflows (e.g.,
\citealt{scoville1986,kraus2006,qiu2011,beuther2012c,sandell2020}),
finding high-velocity emission in its vicinity is not a surprise.
Furthermore, the blue-shifted outflow lobe emanating from IRS1 towards
the north is aligned right with the central area of the H{\sc ii}
region and may explain part of the blue-shifted emission
\citep{sandell2020}.  Nevertheless.  the [CII] emission below
$-60$\,km\,s$^{-1}$ is significantly stronger than the CO(3--2)
emission at the same velocities (Fig.\,\ref{spectra}).  Hence, a sole
explanation by the outflow appears unlikely, and a significant
fraction of the highly blue-shifted gas appears to be associated with
the suggested bubble-structures.

Using the same assumptions as described in the previous section, we
can estimate the mass of ionized carbon associated with this
blue-shifted [CII] component. Integrating the blue-shifted [CII] line
between $-75$ and $-65$\,km\,s$^{-1}$ over the entire emission area
($\sim$22.5\,pc$^2$), we find in the temperature regime between 100
and 150\,K a mass of $M_{\rm{[CII]}}^{\rm{blue}}\approx 0.66\pm
0.1$\,M$_{\odot}$. Although the area is roughly a factor two larger
than that used for the mass estimates for the bar in the previous
section, the mass in this blue-shifted component is more than a
factor 10 smaller than that in the bar-like structure. Nevertheless,
these data show that still a significant fraction of ionized carbon is
found in that mainly in the [CII] line emitting velocity component.

Assuming that at least part of the blue-shifted high-velocity gas
associated with the proposed bubble-like structures is indeed
associated with expanding gas along the line of sight, the fact that
we see barely any atomic or molecular carbon counterpart implies that
there is very low or even no interaction of the expanding gas with a
denser environmental cloud.

\subsection{Shell structures associated with H{\sc ii} regions and
  PDRs in [CII] emission}

As mentioned in the Introduction, the existence of expanding
shell-like structures around H{\sc ii} regions and PDRs is a recurring
result of the SOFIA FEEDBACK program, where NGC7538 is part of, as
well as other [CII] studies. The best studied examples are Orion
\citep{pabst2019}, RCW120 \citep{luisi2021}, and RCW49
\citep{tiwari2021}. From a geometry point of view, the Orion Veil and
RCW120 exhibit more complete shell-like structures in the [CII]
emission. The NGC7538 data presented here also show bubble-like
features, but with much more sub-structure and no real large-scale
shell. RCW49 again exhibits a large-scale shell-structure, however,
depending on the velocities of the gas, the shell is broken open
allowing radiation to escape into the environment.

Regarding the expansion speeds measured in the [CII] emission, all
four regions discussed here have similar expansion velocities between
roughly 10 and 16\,km\,s$^{-1}$. Assuming constant expansion, the
given velocities result in estimated expansion time-scales between 0.1
and 0.27\,Myr, all within a similar range.

The expansion of the four regions can in neither case be explained by
thermal expansion like in the classical Spitzer picture
\citep{spitzer1998}, but wind-driving from the central O-stars is
needed.

Furthermore, it is interesting to note that in three of the regions --
RCW120, RCW49 and NGC7538 -- star formation at the edges of the
expanding regions is observed. For NGC7538, triggering as a potential
scenario is discussed in detail by \citet{sharma2017}. Hence,
triggered star formation seems a regular outcome in such kind of
regions (see also general bubble/protostar correlation analysis by,
e.g., \citealt{kendrew2016} or \citealt{palmeirim2017}). In contrast
to that, no obvious star formation is identified at the edges of the
Orion Veil. \citet{goicoechea2020} recently found small globules of
molecular gas within the Veil structure, however, these are rather
transient objects and unlikely to form stars. Hence, while the star
formation activity around RCW120, RCW49 and NGC7538 hint towards at
least partly positive feedback triggering new star formation events,
in the Orion Veil, the feedback processes expel and reprocess the gas,
limiting further star formation.

\section{Conclusions and Summary}
\label{conclusion}

Using SOFIA, we are able to map roughly 210 square arcmin
(125\,pc$^2$) around the H{\sc ii} region NGC7538 in the
fine-structure line of ionized carbon [CII] as well as fine-structure
[CI] and CO(8--7) emission. These SOFIA data are complemented with
archival near-/far-infrared/cm continuum and CO(3--2) line
observations.

The [CII] emission reveals a wealth of information about the structure
and composition of this region of high-mass star formation. While the
overall [CII] morphology may at first sight appear similar to that of
the ionized gas traced by the cm continuum emission, the
velocity-resolved [CII] emission reveals considerable sub-structure
resembling the PDR otherwise well traced in the 8 and 70\,$\mu$m
emission. Using a automated bubble/ring-identification algorithm, we
identify four bubble-like structures. Two of them show kinematic
signatures of expanding bubbles. One of these structures is centered
near the main exciting sources IRS6 and IRS5. For the other three
bubble-like structures no clear central source can be identified. This
may indicate that at least some of the bubble-like structures could be
mimicked by the expansion of the gas into an intrinsically porous and
inhomogeneous medium within the H{\sc ii} region.

An analysis of the expansion velocities based on high-velocity [CII]
gas components reveals that purely thermal expansion from a
Spitzer-like H{\sc ii} region is not sufficient, but that wind-driving
by the central O-stars is needed. This is supported by Chandra
observations revealing the presence of diffuse X-ray emission due to
hot plasma, resulting from the shocked stellar winds from these O
stars.

The most blue-shifted [CII] velocity component barely shows any
counterpart in atomic or molecular carbon emission.
At the interface of the H{\sc ii} region and the molecular cloud we
find a bar-like photon-dominated region where ionized C$^+$, atomic
C$^0$ and molecular carbon CO reveal a layered structured. The carbon
in the PDR is dominantly in its ionized C$^+$ form. A
velocity-gradient across that bar-like PDR is identified, indicating
that the PDR is pushed along the line of sight in the direction of the
observer.

The [CII] study of this prototypical H{\sc ii} region NGC7538 reveals
a highly sub-structured H{\sc ii} region that interacts with the
adjacent molecular cloud, potentially triggering new star formation
events in some of the adjacent dense cores. Setting the region into
context with other recently investigated [CII] emission studies from
the FEEDBACK program (Orion Veil, RCW120, RCW49), one finds that
bubble-like expanding morphologies are recurring structures in
PDRs. While in some of these regions star formation occurs at the
edges of the expanding regions, indicating positive feedback, the data
from the Orion Veil reveal the destructive forces of feedback,
expelling and reshaping the gas in that region. Hence, both positive
and negative feedback can occur by the expansion of H{\sc ii}
regions. The SOFIA legacy program FEEDBACK will dissect more H{\sc ii}
regions and reveal common processes in H{\sc ii} region-molecular
cloud interaction zones.

\begin{acknowledgements} 
  Financial support for the SOFIA Legacy Program, FEEDBACK, at the
  University of Maryland was provided by NASA through award SOF070077
  issued by USRA. The FEEDBACK project is supported by the BMWI via
  DLR, Projekt Number 50 OR 1916 (FEEDBACK) and Projekt Number 50 OR
  1714 (MOBS - MOdellierung von Beobachtungsdaten SOFIA). The
  Digitized Sky Surveys were produced at the Space Telescope Science
  Institute under U.S. Government grant NAG W-2166. The images of
  these surveys are based on photographic data obtained using the
  Oschin Schmidt Telescope on Palomar Mountain and the UK Schmidt
  Telescope. The plates were processed into the present compressed
  digital form with the permission of these institutions. This work is
  based in part on observations made with the Spitzer Space Telescope,
  which was operated by the Jet Propulsion Laboratory, California
  Institute of Technology under a contract with NASA. We like to
    thank the referee for a constructive report improving the paper.
  HB acknowledges support from the European Research Council under the
  Horizon 2020 Framework Program via the ERC Consolidator Grant
  CSF-648505.  HB also acknowledges support from the Deutsche
  Forschungsgemeinschaft in the Collaborative Research Center SFB 881
  - Project-ID 138713538 - “The Milky Way System” (subproject
  B1). N.S. acknowledges support by the Agence National de Recherche
  (ANR/France) and the Deutsche Forschungsgemeinschaft (DFG/Germany)
  through the project "GENESIS"
  (ANR-16-CE92-0035-01/DFG1591/2-1). This work was supported by the
  German Deutsche Forschungsgemeinschaft, DFG project number SFB
  956. This project has received funding from the European Research
  Council (ERC) under the European Union’s Horizon 2020 research and
  innovation programme (Grant agreement No. 851435).
\end{acknowledgements}

\appendix

\section{Complementary figures}

\begin{figure*}[htb]
\includegraphics[width=0.99\textwidth]{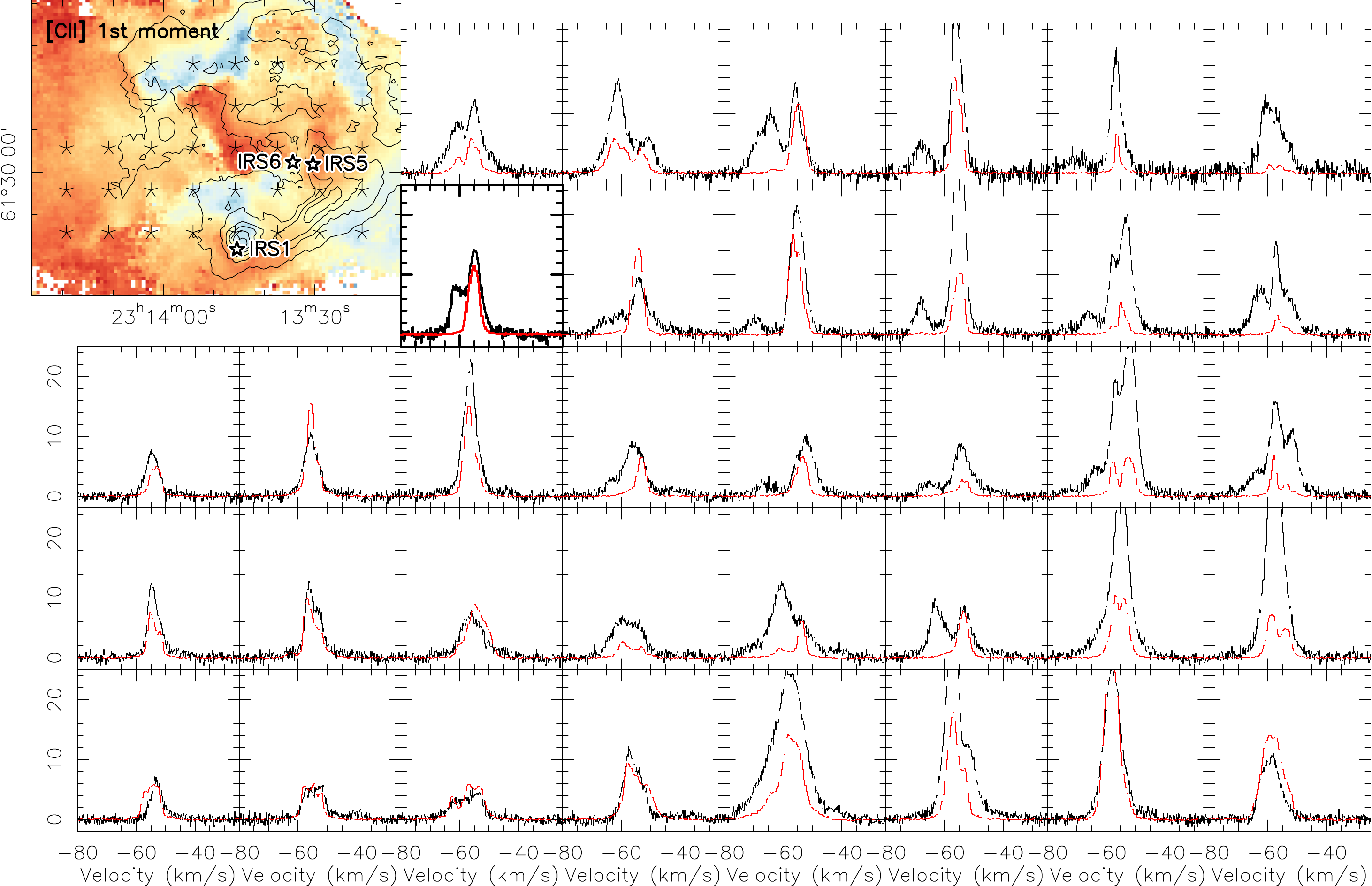}
\caption{[CII] (black) and CO(3--2) spectra towards the grid marked in
  the top-left [CII] 1st moment map. The contours show the [CII]
  emission at only a few selected contour levels to highlight the main
  structures. The sources IRS5, IRS6 and IRS1 are also shown as
  black-and-white stars.}
  \label{spectra_grid} 
\end{figure*} 

\begin{figure*}[htb]
\includegraphics[width=0.99\textwidth]{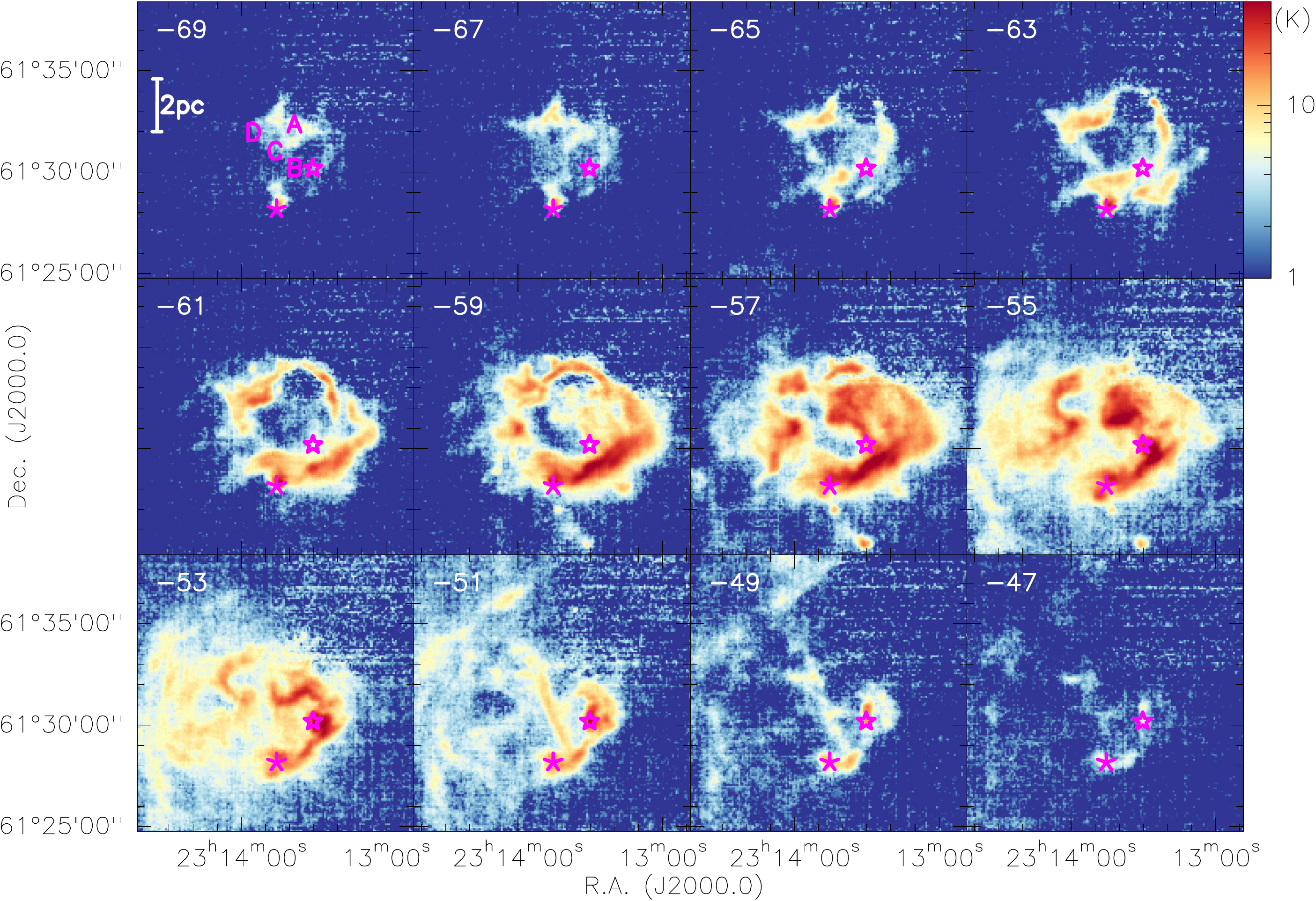}
\caption{[CII] channel map for the velocities labeled in each
  panel. The data are binned for that plot in 2\,km\,s$^{-1}$
  channels. The two purple stars mark the positions of IRS1 and IRS5,
  and a scale-bar is shown in the top-left panel. The intensities are
  always plotted logarithmically from 1 to 40\,K.}
\label{channel_cii_nobubbles} 
\end{figure*} 

\begin{figure*}[htb]
\includegraphics[width=0.99\textwidth]{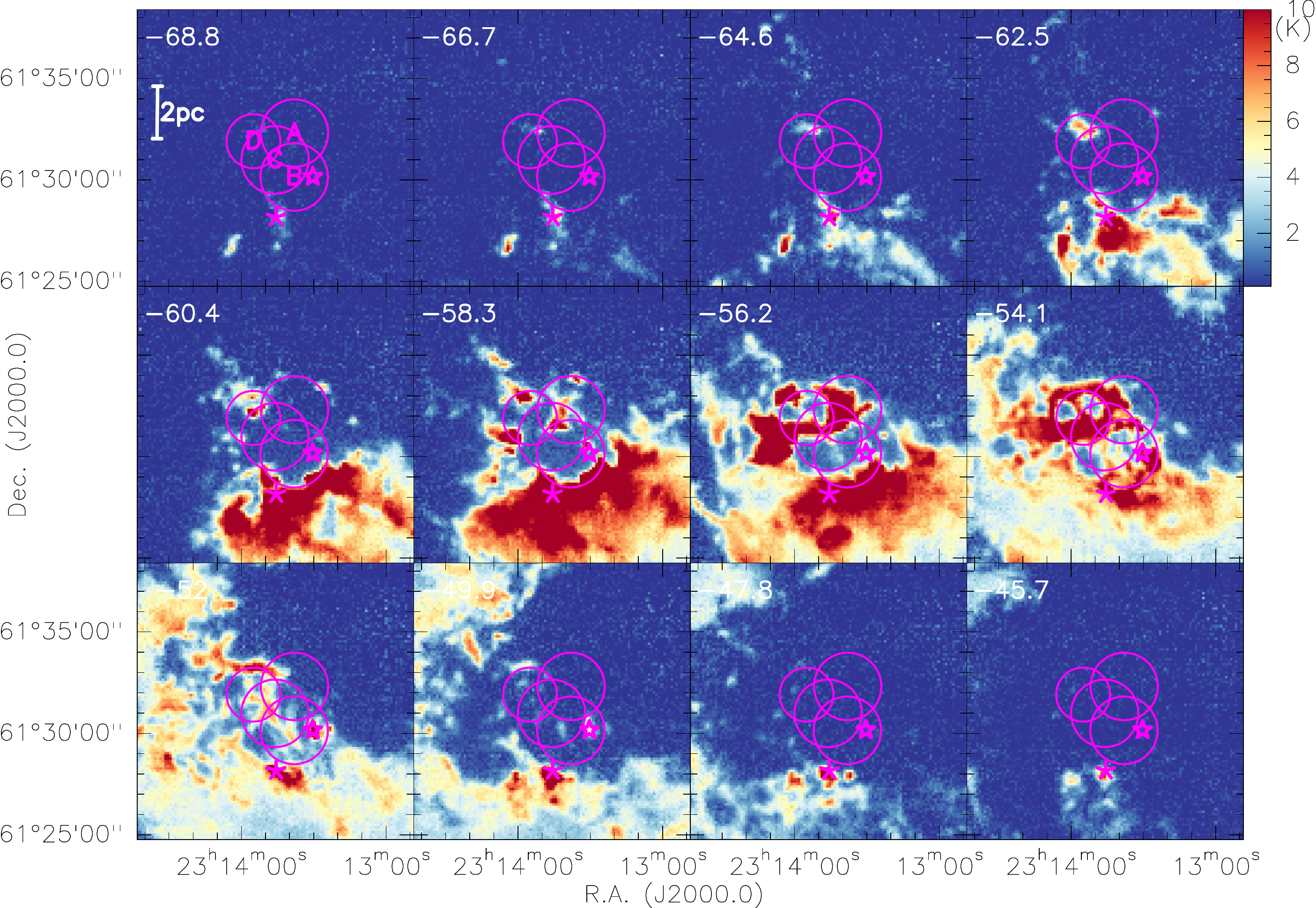}
\caption{CO(3--2) channel map for the velocities labeled in each
  panel. The data are plotted in steps of 2.1\,km\,s$^{-1}$ (5
  channels). The two purple stars mark the positions of IRS1 and IRS5,
  and a scale-bar is shown in the top-left panel. The intensities are
  always plotted linearly from 0.1 to 10\,K. The purple circles
  outline potential bubbles. These bubbles are labeled in the
  top-left panel.}
\label{channel_co} 
\end{figure*}

\begin{figure*}[htb]
\includegraphics[width=0.99\textwidth]{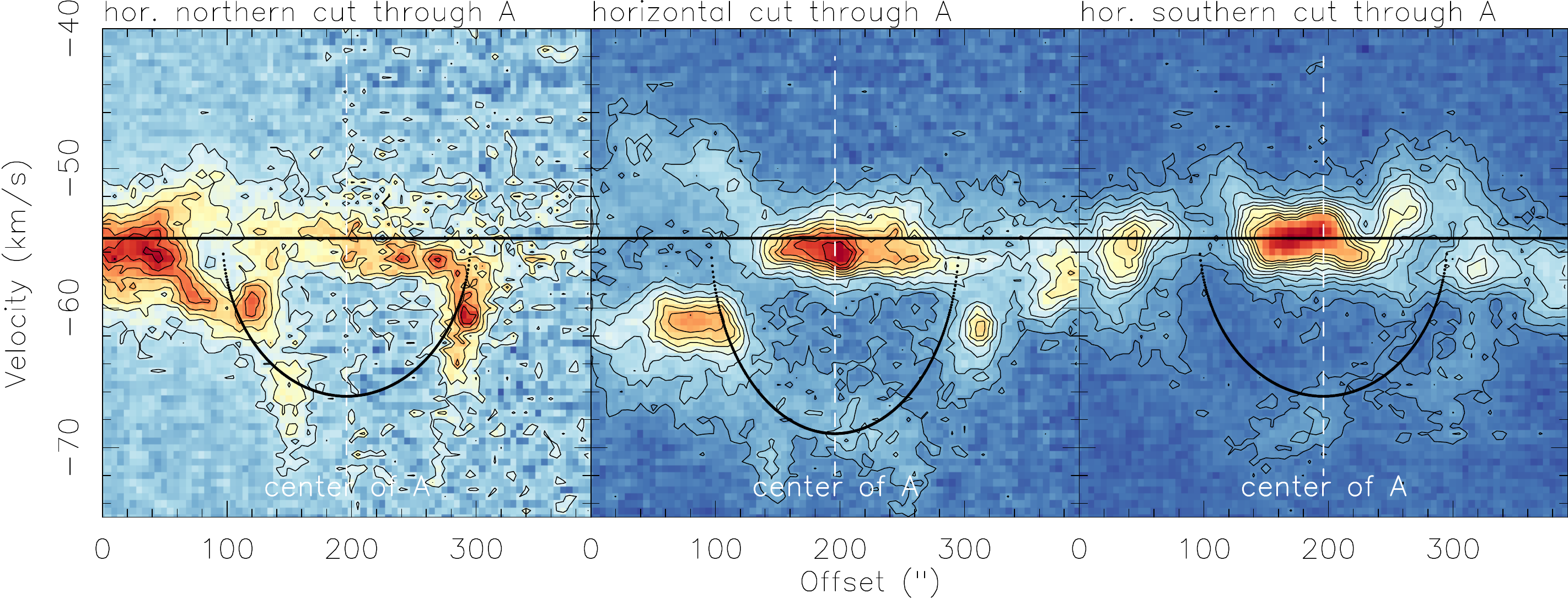}
\caption{Horizontal position-velocity cuts through bubble A from east
  to west derived from the [CII] line emission. The central panel goes
  through the estimated central position of A whereas the left and
  right panels are shifted $1'$ to the north and south,
  respectively. The dashed line marks the approximate center of the
  bubble. The black dotted lines in the middle panel outline the
  elliptical shape a spherical shell with radius $100''$ and expansion
  velocity of 14\,km\,s$^{-1}$ would have in the pv diagram. The
  dotted lines in the left and right panel correspond to the same
  expanding shell at the $1'$ shift to the east and west where the
  $1'$ corresponds to $\sim$36\,deg offset in the sphere (e.g.,
  \citealt{butterfield2018}, sketch in their appendix A.2). Hence the
  velocities are then reduced by $cos(36^{\circ})$.}
\label{pv_bubble_A_hor} 
\end{figure*} 
\clearpage

\section{Bubble finding}
\label{bubble_finding}

The assignment of the individual structures visible in the channel
maps to bubbles and rings with a definite origin can be highly
subjective. To avoid biases of an by-eye-identification, we
implemented a method to quantify the amount of ring-like structures in
individual maps. The goal of the method was to measure structures that
one would identify by eye as rings and quantify their
significance. This approach is complementary to ring-finding
algorithms like CASI \citep{vanoort2019} that allow to scan huge
amounts of data but only provide a binary detection criterion.

To quantify the match between the map and a ring, we compute the
covariance between the map and an annulus structure, that is radially
symmetric and positive between an inner and an outer radius and zero
everywhere else. We normalize the positive value to provide an
integral of unity. The covariance map peaks at the points with the
strongest match between the actual structure and the ring. A search
for the peaks therefore allows for ring finding, but a simple
inspection of the covariance map is insufficient for this because
individual peaks in the intensity distribution translate linearly into
the covariance coefficients. A reliable quantification therefore
requires comparison of the match with different structures. We
perform this through a parameter scan comparing to structures with
different ring radii and widths.

To be consistent with the typical by-eye-identification we limited the
ring width by a ratio between outer and inner radius between 1.1 and
1.6 and scanned this range. However, when applied to the data here, we
found only a weak dependence of the results on this ratio and usually
a good match for ratios around 1.4, so that we limit the further
discussion to rings of this width to reduce the dimensionality of the
problem. Furthermore, we scanned possible ring radii between roughly
38 and 163$''$.

\begin{figure}[ht]
\includegraphics[angle=180,width=0.49\textwidth]{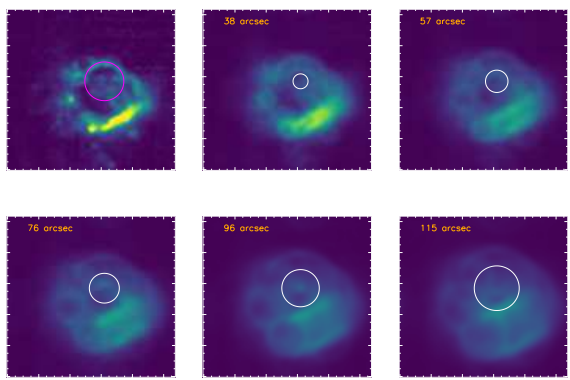}
\caption{Demonstration of the covariance maps for the
          convolution of the $-$61\,km\,s$^{-1}$ channel map (see
          Fig.~\ref{channel_cii}) with rings of different radii. The
          first panel shows the channel map. The magenta circle
          emphasizes one of the obvious ring-like structures with a
          radius of 100$''$. The other five panels show the covariance
          maps from the convolution of rings with increasing
          radii. The radii are indicated at the same position in the
          map.}
        \label{fig_covariance_channel5}
\end{figure}

Fig.~\ref{fig_covariance_channel5} demonstrates this for the [CII]
channel map at $-61$\,km\,s$^{-1}$ from Fig.~\ref{channel_cii}. The
first panel shows the channel map emphasizing one of the prominent
arcs with a radius of $\sim 100''$. The other panels show the
covariance maps from the convolution with rings of 38, 57, 76, 96, and
115$''$. They are indicated as thin circles centered at the same
position. One can see that for small radii the covariance maps show a
dip in the centre of the circles, while only for the 96$''$ radius,
the map shows a peak in the centre. Considering only the location at
the center of the circle and scanning the ring radii we find the
maximum covariance coefficient for the ``correct'' ring
radius. However, it is interesting to see that the covariance map for
76$''$ rings also shows a peak in the vicinity, shifted to the north
by about 20$''$. This indicates that the arc in the channel map could
also be fitted with a smaller ring that is spatially shifted. From
these covariance maps only, neither of the two solutions can be
excluded.

Additional information can be obtained when including the velocity
information if we do not only look for rings, but the physical
scenario of an expanding bubble. In that case, only at the systemic
zero velocity the bubble appears as a ring with its full diameter. At
lower and higher velocities, the channel maps perform a cut through
the bubble in front or behind the central plane so that a smaller
broadened ring appears. A different way to visualize this is via
plotting the covariance coefficients against the velocity and the ring
radius. For an ideal bubble, one then finds an inverse ``V'' shape
where the largest ring radius is detected at zero systemic
velocities. At intermediate velocities one finds a smaller ring.

To look for bubbles in the NGC7538 [CII] data we can look for this
inverse ``V'' shape in the covariance coefficients computed for all
channel maps and ring radii.
We perform a pre-selection of pixels by looking for rings in the
individual channel maps. The principle is straight forward. We search
for peaks on the covariance maps that are small, isotropic and
sufficiently bright (for details see Kong et al.~in prep.). Here, we
perform a GAUCSCLUMPS \citep{stutzki1990} decomposition of the
covariance maps and locate clumps that have an axis ratio of less than
two, a radius below the inner ring radius of the convolution and an
intensity of more than 5~K in the [CII] channel maps.

\begin{figure}[ht]
\includegraphics[angle=180,width=0.49\textwidth]{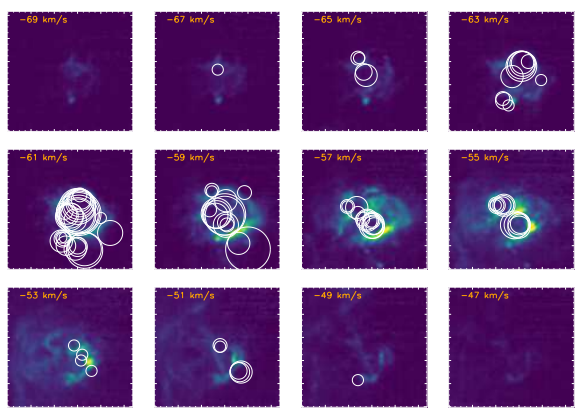}
	\caption{[CII] channel map with an overlay of all ring
          candidates identified in the individual channels. We plotted
          the rings not with their full width used in the convolution
          (outer radius 1.4 times the inner radius) but only as very
          thin rings to allow for a distinction of the different radii
          by eye.}\label{fig_channel-ring-candidates}
\end{figure}

The result is shown in Fig.~\ref{fig_channel-ring-candidates} for all
[CII] channel maps. The number of ring candidates varies strongly
between the different channels. At the extreme velocities, no rings
are found, but the largest number of rings is not found at the
velocities of the brightest intensity but in panel 5, for a velocity
of $-61$\,km\,s$^{-1}$. In many cases the radius of the ring is not
accurately confined. We find overlapping rings with nearby
centres. This is in particular the case for panel 5
($-61$\,km\,s$^{-1}$) where five different radii can fit the central
ring structure. A typical effect is best visible in panel 10, at
$-51$\,km\,s$^{-1}$, where an arc like structure can be fitted by
rings of different radii and simultaneously different centres. This
demonstrates the fundamental uncertainty of the approach. Based on the
plots here, it is only possible to measure the location of the center
of each ring if we a priori know its radius. Fitting both leaves some
uncertainty.

The locations of all these rings gives a guidance where to look
for the inverse ``V'' structure in the four-dimensional covariance
cube. If we look at the distribution of the different ring centres of
the different channel maps (Fig.~\ref{fig_channel-ring-candidates}) we
find three dominant clusters that involve rings of varying radii\\
(A) R.A.(J2000.0)=23h13m38s, Dec=61$^{\circ}$32$'$20$''$\\
(B) R.A.(J2000.0)=23h13m38s, Dec=61$^{\circ}$30$'$12$''$\\
(C) R.A.(J2000.0)=23h13m46s, Dec=61$^{\circ}$31$'$00$''$\\
A cluster with only small ring radii is located around\\
(D) R.A.(J2000.0)=23h13m55s, Dec=61$^{\circ}$31$'$56$''$\\
and a cluster with only large radii around\\
(E) R.A.(J2000.0)=23h13m39s, Dec=61$^{\circ}$31$'$16$''$.\\
From the inspection of Fig.~\ref{fig_channel-ring-candidates} we can
exclude most other positions because they only detected some outer
arcs, not driven from inside of the region. In addition to that, the
largest ring E appears as an overlap of the other smaller rings, in
particular A, B and C (Figs.~\ref{fig_channel-ring-candidates} \&
\ref{fig_point-rings}). Therefore, although identified by the
algorithm, we do not consider E as a separate physical entity.

\begin{figure}[htb]
\includegraphics[angle=180,width=0.49\textwidth]{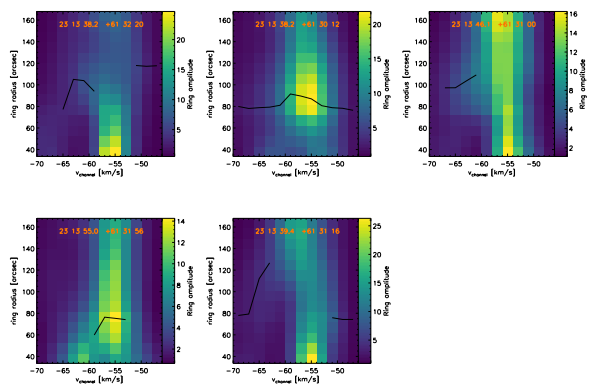}
	\caption{Covariance coefficients at the positions of the five
          prominent clusters from
          Fig.~\ref{fig_channel-ring-candidates}. The coefficients are
          shown as a function of the channel velocity and the ring
          radius used in the convolution. The black lines mark the
          maxima of the covariance coefficients in terms of the ring
          radius. In those channels without black line, no ring could
          be identified.}
	\label{fig_point-analysis}
\end{figure}

Fig.~\ref{fig_point-analysis} shows the covariance coefficients for
these five locations as a function of channel velocity and ring
radius. The relevant information is the location and width of the peak
in the coefficients as a function of the ring radius, in particular
the shape of the line formed by those peaks as a function of the
velocity channel.  All panels show somewhat different characteristics
but also show at least some indication of the inverse ``V'' shape.
Fig.~\ref{fig_point-analysis} shows as a black line the maxima of the
covariance coefficients that can be considered as a proxy for the
ring/bubble radius. The width of these maxima can be considered as an
estimator on the uncertainty on the ring radius. The ``V'' shape is
most pronounced for structure D (bottom-left panel in
Fig.~\ref{fig_point-analysis}). There the maximum ring radius is
$\sim$80$''$ at velocities of $-57$ and $-55$\,km\,s$^{-1}$ and we
find dropping radii to higher and lower velocities. For the other
bubble candidates, the transition to smaller radii and lower velocity
is less well determined. In the following, we use as approximate radii
for structures A, B and C $100''$, and for structure D $80''$.

\begin{figure}[ht]
\includegraphics[angle=180,width=0.49\textwidth]{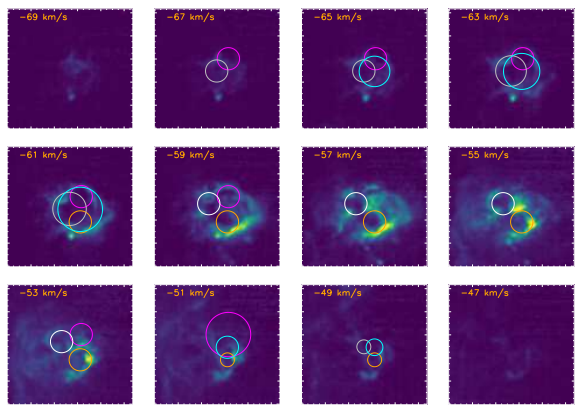}
	\caption{[CII] channel map with an overlay of the identified
          rings around the five candidate positions from
          Fig.~\ref{fig_point-analysis}. The ring radii vary with
          velocity. The plotted rings are narrower than the rings used
          in the convolution. Different ring positions are encoded in
          a different ring color. The positions in panels 1--5 from
          Fig.~\ref{fig_point-analysis} correspond to the colors
          magenta (A), orange (B), grey (C), white (D), and cyan (E).}
	\label{fig_point-rings}
\end{figure}

Figure~\ref{fig_point-rings} shows the same result in a different
way. Here we plot again the channel map from Fig.~\ref{channel_cii}
but now over-plot circles giving the mean ring diameter for the ring
that is the local peak in the covariance map for the five points
discussed in Fig.~\ref{fig_point-analysis}. This means for each
channel we select the corresponding velocity column in the covariance
plots in Fig.~\ref{fig_point-analysis}, look for the maximum and draw
the corresponding circle if the maximum is not at the minimum radius
and above 4\,\% of the peak intensity. The five positions are color
coded. The systematic behaviour of an increasing ring diameter with
velocity is well visible for the grey and cyan rings representing
position C and E. They behave like the ideal bubble at least for
blue-shifted velocities. In contrast the magenta and white rings,
positions A and D, are relatively constant in diameter. They rather
represent ring-structures in ppv-space. The matches at a velocity of
$-51$\,km\,s$^{-1}$ rather look, at least partially, like a
coincidence, not necessarily connected to the rings or bubbles at
lower velocities. This all indicates that a bubble expansion in the
region was hindered towards red-shifted velocities. It happens mainly
at blue-shifted velocities, this means towards the observer.

To estimate the accuracy of the determination of the ring positions
and radii, we investigated the covariance plots for varying central
position of the identified bubbles. This analysis indicates that for
the conditions in NGC7538 with ring or bubble radii on the order of
60--160$''$ we can pin down their origin with an accuracy of
20--30$''$.  The bubble diameter should be the largest one found in
the central velocity channel. If we consider the rings from Fig.~\ref
{fig_point-rings} and compare the corresponding ring candidates in
Fig.~\ref{fig_channel-ring-candidates} we see that in those velocity
channels typically two or three large rings appear corresponding to an
uncertainty of the ring radius of 20--30$''$ (see also lines in
Fig.~\ref{fig_point-analysis}).


\begin{thebibliography}{73}
\expandafter\ifx\csname natexlab\endcsname\relax\def\natexlab#1{#1}\fi

\bibitem[{{Beuther} {et~al.}(2012){Beuther}, {Linz}, \&
  {Henning}}]{beuther2012c}
{Beuther}, H., {Linz}, H., \& {Henning}, T. 2012, \aap, 543, A88

\bibitem[{{Beuther} {et~al.}(2018){Beuther}, {Mottram}, {Ahmadi}, {Bosco},
  {Linz}, {Henning}, {Klaassen}, {Winters}, {Maud}, {Kuiper}, {Semenov},
  {Gieser}, {Peters}, {Urquhart}, {Pudritz}, {Ragan}, {Feng}, {Keto},
  {Leurini}, {Cesaroni}, {Beltran}, {Palau}, {S{\'a}nchez-Monge},
  {Galvan-Madrid}, {Zhang}, {Schilke}, {Wyrowski}, {Johnston}, {Longmore},
  {Lumsden}, {Hoare}, {Menten}, \& {Csengeri}}]{beuther2018b}
{Beuther}, H., {Mottram}, J.~C., {Ahmadi}, A., {et~al.} 2018, \aap, 617, A100

\bibitem[{{Beuther} {et~al.}(2014){Beuther}, {Ragan}, {Ossenkopf}, {Glover},
  {Henning}, {Linz}, {Nielbock}, {Krause}, {Stutzki}, {Schilke}, \&
  {G{\"u}sten}}]{beuther2014}
{Beuther}, H., {Ragan}, S.~E., {Ossenkopf}, V., {et~al.} 2014, \aap, 571, A53

\bibitem[{{Butterfield} {et~al.}(2018){Butterfield}, {Lang}, {Morris}, {Mills},
  \& {Ott}}]{butterfield2018}
{Butterfield}, N., {Lang}, C.~C., {Morris}, M., {Mills}, E. A.~C., \& {Ott}, J.
  2018, \apj, 852, 11

\bibitem[{{Cabrit} {et~al.}(1988){Cabrit}, {Goldsmith}, \&
  {Snell}}]{cabrit1988}
{Cabrit}, S., {Goldsmith}, P.~F., \& {Snell}, R.~L. 1988, \apj, 334, 196

\bibitem[{{Drew} {et~al.}(2005){Drew}, {Greimel}, {Irwin}, {Aungwerojwit},
  {Barlow}, {Corradi}, {Drake}, {G{\"a}nsicke}, {Groot}, {Hales}, {Hopewell},
  {Irwin}, {Knigge}, {Leisy}, {Lennon}, {Mampaso}, {Masheder}, {Matsuura},
  {Morales-Rueda}, {Morris}, {Parker}, {Phillipps}, {Rodriguez-Gil}, {Roelofs},
  {Skillen}, {Sokoloski}, {Steeghs}, {Unruh}, {Viironen}, {Vink}, {Walton},
  {Witham}, {Wright}, {Zijlstra}, \& {Zurita}}]{drew2005}
{Drew}, J.~E., {Greimel}, R., {Irwin}, M.~J., {et~al.} 2005, \mnras, 362, 753

\bibitem[{{Elmegreen}(2011)}]{elmegreen2011b}
{Elmegreen}, B.~G. 2011, in EAS Publications Series, Vol.~51, EAS Publications
  Series, ed. C.~{Charbonnel} \& T.~{Montmerle}, 45--58

\bibitem[{{Fallscheer} {et~al.}(2013){Fallscheer}, {Reid}, {Di Francesco},
  {Martin}, {Hill}, {Hennemann}, {Nguyen-Luong}, {Motte}, {Men'shchikov},
  {Andr{\'e}}, {Ward-Thompson}, {Griffin}, {Kirk}, {Konyves}, {Rygl},
  {Sadavoy}, {Sauvage}, {Schneider}, {Anderson}, {Benedettini}, {Bernard},
  {Bontemps}, {Ginsburg}, {Molinari}, {Polychroni}, {Rivera-Ingraham},
  {Roussel}, {Testi}, {White}, {Williams}, {Wilson}, {Wong}, \&
  {Zavagno}}]{fallscheer2013}
{Fallscheer}, C., {Reid}, M.~A., {Di Francesco}, J., {et~al.} 2013, \apj, 773,
  102

\bibitem[{{Fazio} {et~al.}(2004){Fazio}, {Hora}, {Allen}, {Ashby}, {Barmby},
  {Deutsch}, {Huang}, {Kleiner}, {Marengo}, {Megeath}, {Melnick}, {Pahre},
  {Patten}, {Polizotti}, {Smith}, {Taylor}, {Wang}, {Willner}, {Hoffmann},
  {Pipher}, {Forrest}, {McMurty}, {McCreight}, {McKelvey}, {McMurray}, {Koch},
  {Moseley}, {Arendt}, {Mentzell}, {Marx}, {Losch}, {Mayman}, {Eichhorn},
  {Krebs}, {Jhabvala}, {Gezari}, {Fixsen}, {Flores}, {Shakoorzadeh}, {Jungo},
  {Hakun}, {Workman}, {Karpati}, {Kichak}, {Whitley}, {Mann}, {Tollestrup},
  {Eisenhardt}, {Stern}, {Gorjian}, {Bhattacharya}, {Carey}, {Nelson},
  {Glaccum}, {Lacy}, {Lowrance}, {Laine}, {Reach}, {Stauffer}, {Surace},
  {Wilson}, {Wright}, {Hoffman}, {Domingo}, \& {Cohen}}]{fazio2004}
{Fazio}, G.~G., {Hora}, J.~L., {Allen}, L.~E., {et~al.} 2004, \apjs, 154, 10

\bibitem[{{Frerking} {et~al.}(1989){Frerking}, {Keene}, {Blake}, \&
  {Phillips}}]{frerking1989}
{Frerking}, M.~A., {Keene}, J., {Blake}, G.~A., \& {Phillips}, T.~G. 1989,
  \apj, 344, 311

\bibitem[{{Geen} {et~al.}(2016){Geen}, {Hennebelle}, {Tremblin}, \&
  {Rosdahl}}]{geen2016}
{Geen}, S., {Hennebelle}, P., {Tremblin}, P., \& {Rosdahl}, J. 2016, \mnras,
  463, 3129

\bibitem[{{Geen} {et~al.}(2018){Geen}, {Watson}, {Rosdahl}, {Bieri}, {Klessen},
  \& {Hennebelle}}]{geen2018}
{Geen}, S., {Watson}, S.~K., {Rosdahl}, J., {et~al.} 2018, \mnras, 481, 2548

\bibitem[{{Gerlich}(1990)}]{gerlich1990}
{Gerlich}, D. 1990, \jcp, 92, 2377

\bibitem[{{Gibson} {et~al.}(2005){Gibson}, {Taylor}, {Higgs}, {Brunt}, \&
  {Dewdney}}]{gibson2005b}
{Gibson}, S.~J., {Taylor}, A.~R., {Higgs}, L.~A., {Brunt}, C.~M., \& {Dewdney},
  P.~E. 2005, \apj, 626, 214

\bibitem[{{Goicoechea} {et~al.}(2020){Goicoechea}, {Pabst}, {Kabanovic},
  {Santa-Maria}, {Marcelino}, {Tielens}, {Hacar}, {Bern{\'e}}, {Buchbender},
  {Cuadrado}, {Higgins}, {Kramer}, {Stutzki}, {Suri}, {Teyssier}, \&
  {Wolfire}}]{goicoechea2020}
{Goicoechea}, J.~R., {Pabst}, C.~H.~M., {Kabanovic}, S., {et~al.} 2020, \aap,
  639, A1

\bibitem[{{Goldsmith} {et~al.}(2012){Goldsmith}, {Langer}, {Pineda}, \&
  {Velusamy}}]{goldsmith2012}
{Goldsmith}, P.~F., {Langer}, W.~D., {Pineda}, J.~L., \& {Velusamy}, T. 2012,
  \apjs, 203, 13

\bibitem[{{Guan} {et~al.}(2012){Guan}, {Stutzki}, {Graf}, {G{\"u}sten},
  {Okada}, {Requena-Torres}, {Simon}, \& {Wiesemeyer}}]{guan2012}
{Guan}, X., {Stutzki}, J., {Graf}, U.~U., {et~al.} 2012, \aap, 542, L4

\bibitem[{{G{\"u}del} {et~al.}(2008){G{\"u}del}, {Briggs}, {Montmerle},
  {Audard}, {Rebull}, \& {Skinner}}]{guedel2008}
{G{\"u}del}, M., {Briggs}, K.~R., {Montmerle}, T., {et~al.} 2008, Science, 319,
  309

\bibitem[{{Henshaw} {et~al.}(2021){Henshaw}, {Krumholz}, {Butterfield},
  {Mackey}, {Ginsburg}, {Haworth}, {Nogueras-Lara}, {Barnes}, {Longmore},
  {Bally}, {Kruijssen}, {Mills}, {Beuther}, {Walker}, {Battersby}, {Bulatek},
  {Henning}, {Ott}, \& {Soler}}]{henshaw2021}
{Henshaw}, J.~D., {Krumholz}, M.~R., {Butterfield}, N.~O., {et~al.} 2021, arXiv
  e-prints, arXiv:2110.11367

\bibitem[{{Hollenbach} \& {Tielens}(1997)}]{hollenbach1997}
{Hollenbach}, D.~J. \& {Tielens}, A.~G.~G.~M. 1997, \araa, 35, 179

\bibitem[{{Hopkins} {et~al.}(2014){Hopkins}, {Kere{\v{s}}}, {O{\~n}orbe},
  {Faucher-Gigu{\`e}re}, {Quataert}, {Murray}, \& {Bullock}}]{hopkins2014}
{Hopkins}, P.~F., {Kere{\v{s}}}, D., {O{\~n}orbe}, J., {et~al.} 2014, \mnras,
  445, 581

\bibitem[{{Kendrew} {et~al.}(2016){Kendrew}, {Beuther}, {Simpson}, {Csengeri},
  {Wienen}, {Lintott}, {Povich}, {Beaumont}, \& {Schuller}}]{kendrew2016}
{Kendrew}, S., {Beuther}, H., {Simpson}, R., {et~al.} 2016, \apj, 825, 142

\bibitem[{{Kim} {et~al.}(2018){Kim}, {Kim}, \& {Ostriker}}]{kim2018}
{Kim}, J.-G., {Kim}, W.-T., \& {Ostriker}, E.~C. 2018, \apj, 859, 68

\bibitem[{{Klein} {et~al.}(2012){Klein}, {Hochg{\"u}rtel}, {Kr{\"a}mer},
  {Bell}, {Meyer}, \& {G{\"u}sten}}]{klein2012}
{Klein}, B., {Hochg{\"u}rtel}, S., {Kr{\"a}mer}, I., {et~al.} 2012, \aap, 542,
  L3

\bibitem[{{Kraus} {et~al.}(2006){Kraus}, {Balega}, {Elitzur}, {Hofmann},
  {Preibisch}, {Rosen}, {Schertl}, {Weigelt}, \& {Young}}]{kraus2006}
{Kraus}, S., {Balega}, Y., {Elitzur}, M., {et~al.} 2006, \aap, 455, 521

\bibitem[{{Lancaster} {et~al.}(2021{\natexlab{a}}){Lancaster}, {Ostriker},
  {Kim}, \& {Kim}}]{lancaster2021a}
{Lancaster}, L., {Ostriker}, E.~C., {Kim}, J.-G., \& {Kim}, C.-G.
  2021{\natexlab{a}}, \apj, 914, 89

\bibitem[{{Lancaster} {et~al.}(2021{\natexlab{b}}){Lancaster}, {Ostriker},
  {Kim}, \& {Kim}}]{lancaster2021b}
{Lancaster}, L., {Ostriker}, E.~C., {Kim}, J.-G., \& {Kim}, C.-G.
  2021{\natexlab{b}}, \apj, 914, 90

\bibitem[{{Lancaster} {et~al.}(2021{\natexlab{c}}){Lancaster}, {Ostriker},
  {Kim}, \& {Kim}}]{lancaster2021c}
{Lancaster}, L., {Ostriker}, E.~C., {Kim}, J.-G., \& {Kim}, C.-G.
  2021{\natexlab{c}}, arXiv e-prints, arXiv:2110.05508

\bibitem[{{Langer} {et~al.}(2010){Langer}, {Velusamy}, {Pineda}, {Goldsmith},
  {Li}, \& {Yorke}}]{langer2010}
{Langer}, W.~D., {Velusamy}, T., {Pineda}, J.~L., {et~al.} 2010, \aap, 521, L17

\bibitem[{{Le Bourlot}(1991)}]{lebourlot1991}
{Le Bourlot}, J. 1991, \aap, 242, 235

\bibitem[{{Lique} {et~al.}(2013){Lique}, {Werfelli}, {Halvick}, {Stoecklin},
  {Faure}, {Wiesenfeld}, \& {Dagdigian}}]{lique2013}
{Lique}, F., {Werfelli}, G., {Halvick}, P., {et~al.} 2013, \jcp, 138, 204314

\bibitem[{{Luisi} {et~al.}(2016){Luisi}, {Anderson}, {Balser}, {Bania}, \&
  {Wenger}}]{luisi2016}
{Luisi}, M., {Anderson}, L.~D., {Balser}, D.~S., {Bania}, T.~M., \& {Wenger},
  T.~V. 2016, \apj, 824, 125

\bibitem[{{Luisi} {et~al.}(2021){Luisi}, {Anderson}, {Schneider}, {Simon},
  {Kabanovic}, {G{\"u}sten}, {Zavagno}, {Broos}, {Buchbender}, {Guevara},
  {Jacobs}, {Justen}, {Klein}, {Linville}, {R{\"o}llig}, {Russeil}, {Stutzki},
  {Tiwari}, {Townsley}, \& {Tielens}}]{luisi2021}
{Luisi}, M., {Anderson}, L.~D., {Schneider}, N., {et~al.} 2021, Science
  Advances, 7, eabe9511

\bibitem[{{Mac Low} \& {McCray}(1988)}]{maclow1988}
{Mac Low}, M.-M. \& {McCray}, R. 1988, \apj, 324, 776

\bibitem[{{Matzner}(2002)}]{matzner2002}
{Matzner}, C.~D. 2002, \apj, 566, 302

\bibitem[{{Moscadelli} {et~al.}(2009){Moscadelli}, {Reid}, {Menten},
  {Brunthaler}, {Zheng}, \& {Xu}}]{moscadelli2009}
{Moscadelli}, L., {Reid}, M.~J., {Menten}, K.~M., {et~al.} 2009, \apj, 693, 406

\bibitem[{{Motte} {et~al.}(2010){Motte}, {Zavagno}, {Bontemps}, {Schneider},
  {Hennemann}, {di Francesco}, {Andr{\'e}}, {Saraceno}, {Griffin}, {Marston},
  {Ward-Thompson}, {White}, {Minier}, {Men'shchikov}, {Hill}, {Abergel},
  {Anderson}, {Aussel}, {Balog}, {Baluteau}, {Bernard}, {Cox}, {Csengeri},
  {Deharveng}, {Didelon}, {di Giorgio}, {Hargrave}, {Huang}, {Kirk}, {Leeks},
  {Li}, {Martin}, {Molinari}, {Nguyen-Luong}, {Olofsson}, {Persi}, {Peretto},
  {Pezzuto}, {Roussel}, {Russeil}, {Sadavoy}, {Sauvage}, {Sibthorpe},
  {Spinoglio}, {Testi}, {Teyssier}, {Vavrek}, {Wilson}, \&
  {Woodcraft}}]{motte2010}
{Motte}, F., {Zavagno}, A., {Bontemps}, S., {et~al.} 2010, \aap, 518, L77+

\bibitem[{{Muijres} {et~al.}(2012){Muijres}, {Vink}, {de Koter}, {M{\"u}ller},
  \& {Langer}}]{muijres2012}
{Muijres}, L.~E., {Vink}, J.~S., {de Koter}, A., {M{\"u}ller}, P.~E., \&
  {Langer}, N. 2012, \aap, 537, A37

\bibitem[{{Pabst} {et~al.}(2019){Pabst}, {Higgins}, {Goicoechea}, {Teyssier},
  {Berne}, {Chambers}, {Wolfire}, {Suri}, {Guesten}, {Stutzki}, {Graf},
  {Risacher}, \& {Tielens}}]{pabst2019}
{Pabst}, C., {Higgins}, R., {Goicoechea}, J.~R., {et~al.} 2019, \nat, 565, 618

\bibitem[{{Pabst} {et~al.}(2020){Pabst}, {Goicoechea}, {Teyssier}, {Bern{\'e}},
  {Higgins}, {Chambers}, {Kabanovic}, {G{\"u}sten}, {Stutzki}, \&
  {Tielens}}]{pabst2020}
{Pabst}, C.~H.~M., {Goicoechea}, J.~R., {Teyssier}, D., {et~al.} 2020, \aap,
  639, A2

\bibitem[{{Palmeirim} {et~al.}(2013){Palmeirim}, {Andr{\'e}}, {Kirk},
  {Ward-Thompson}, {Arzoumanian}, {K{\"o}nyves}, {Didelon}, {Schneider},
  {Benedettini}, {Bontemps}, {Di Francesco}, {Elia}, {Griffin}, {Hennemann},
  {Hill}, {Martin}, {Men'shchikov}, {Molinari}, {Motte}, {Nguyen Luong},
  {Nutter}, {Peretto}, {Pezzuto}, {Roy}, {Rygl}, {Spinoglio}, \&
  {White}}]{palmeirim2013}
{Palmeirim}, P., {Andr{\'e}}, P., {Kirk}, J., {et~al.} 2013, \aap, 550, A38

\bibitem[{{Palmeirim} {et~al.}(2017){Palmeirim}, {Zavagno}, {Elia}, {Moore},
  {Whitworth}, {Tremblin}, {Traficante}, {Merello}, {Russeil}, {Pezzuto},
  {Cambr{\'e}sy}, {Baldeschi}, {Bandieramonte}, {Becciani}, {Benedettini},
  {Buemi}, {Bufano}, {Bulpitt}, {Butora}, {Carey}, {Costa}, {Deharveng}, {Di
  Giorgio}, {Eden}, {Hajnal}, {Hoare}, {Kacsuk}, {Leto}, {Marsh}, {M{\`e}ge},
  {Molinari}, {Molinaro}, {Noriega-Crespo}, {Schisano}, {Sciacca}, {Trigilio},
  {Umana}, \& {Vitello}}]{palmeirim2017}
{Palmeirim}, P., {Zavagno}, A., {Elia}, D., {et~al.} 2017, \aap, 605, A35

\bibitem[{{P{\'e}rez-Beaupuits}
  {et~al.}(2015{\natexlab{a}}){P{\'e}rez-Beaupuits}, {G{\"u}sten}, {Spaans},
  {Ossenkopf}, {Menten}, {Requena-Torres}, {Wiesemeyer}, {Stutzki}, {Guevara},
  \& {Simon}}]{perez2015b}
{P{\'e}rez-Beaupuits}, J.~P., {G{\"u}sten}, R., {Spaans}, M., {et~al.}
  2015{\natexlab{a}}, \aap, 583, A107

\bibitem[{{P{\'e}rez-Beaupuits}
  {et~al.}(2015{\natexlab{b}}){P{\'e}rez-Beaupuits}, {Stutzki}, {Ossenkopf},
  {Spaans}, {G{\"u}sten}, \& {Wiesemeyer}}]{perez2015}
{P{\'e}rez-Beaupuits}, J.~P., {Stutzki}, J., {Ossenkopf}, V., {et~al.}
  2015{\natexlab{b}}, \aap, 575, A9

\bibitem[{{Pilbratt} {et~al.}(2010){Pilbratt}, {Riedinger}, {Passvogel},
  {Crone}, {Doyle}, {Gageur}, {Heras}, {Jewell}, {Metcalfe}, {Ott}, \&
  {Schmidt}}]{A&ASpecialIssue-HERSCHEL}
{Pilbratt}, G.~L., {Riedinger}, J.~R., {Passvogel}, T., {et~al.} 2010, \aap,
  518, L1

\bibitem[{{Poglitsch} {et~al.}(2010){Poglitsch}, {Waelkens}, {Geis},
  {Feuchtgruber}, {Vandenbussche}, {Rodriguez}, {Krause}, {Renotte}, {van
  Hoof}, {Saraceno}, {Cepa}, {Kerschbaum}, {Agn{\`e}se}, {Ali}, {Altieri},
  {Andreani}, {Augueres}, {Balog}, {Barl}, {Bauer}, {Belbachir}, {Benedettini},
  {Billot}, {Boulade}, {Bischof}, {Blommaert}, {Callut}, {Cara}, {Cerulli},
  {Cesarsky}, {Contursi}, {Creten}, {De Meester}, {Doublier}, {Doumayrou},
  {Duband}, {Exter}, {Genzel}, {Gillis}, {Gr{\"o}zinger}, {Henning},
  {Herreros}, {Huygen}, {Inguscio}, {Jakob}, {Jamar}, {Jean}, {de Jong},
  {Katterloher}, {Kiss}, {Klaas}, {Lemke}, {Lutz}, {Madden}, {Marquet},
  {Martignac}, {Mazy}, {Merken}, {Montfort}, {Morbidelli}, {M{\"u}ller},
  {Nielbock}, {Okumura}, {Orfei}, {Ottensamer}, {Pezzuto}, {Popesso},
  {Putzeys}, {Regibo}, {Reveret}, {Royer}, {Sauvage}, {Schreiber}, {Stegmaier},
  {Schmitt}, {Schubert}, {Sturm}, {Thiel}, {Tofani}, {Vavrek}, {Wetzstein},
  {Wieprecht}, \& {Wiezorrek}}]{A&ASpecialIssue-PACS}
{Poglitsch}, A., {Waelkens}, C., {Geis}, N., {et~al.} 2010, \aap, 518, L2

\bibitem[{{Puga} {et~al.}(2010){Puga}, {Mar{\'{\i}}n-Franch}, {Najarro},
  {Lenorzer}, {Herrero}, {Acosta Pulido}, {Chavarr{\'{\i}}a}, {Bik}, {Figer},
  \& {Ram{\'{\i}}rez Alegr{\'{\i}}a}}]{puga2010}
{Puga}, E., {Mar{\'{\i}}n-Franch}, A., {Najarro}, F., {et~al.} 2010, \aap, 517,
  A2

\bibitem[{{Qiu} {et~al.}(2011){Qiu}, {Zhang}, \& {Menten}}]{qiu2011}
{Qiu}, K., {Zhang}, Q., \& {Menten}, K.~M. 2011, \apj, 728, 6

\bibitem[{{Reid} {et~al.}(2019){Reid}, {Menten}, {Brunthaler}, {Zheng}, {Dame},
  {Xu}, {Li}, {Sakai}, {Wu}, {Immer}, {Zhang}, {Sanna}, {Moscadelli}, {Rygl},
  {Bartkiewicz}, {Hu}, {Quiroga-Nu{\~n}ez}, \& {van Langevelde}}]{reid2019}
{Reid}, M.~J., {Menten}, K.~M., {Brunthaler}, A., {et~al.} 2019, \apj, 885, 131

\bibitem[{{Risacher} {et~al.}(2018){Risacher}, {G{\"u}sten}, {Stutzki},
  {H{\"u}bers}, {Aladro}, {Bell}, {Buchbender}, {B{\"u}chel}, {Csengeri},
  {Duran}, {Graf}, {Higgins}, {Honingh}, {Jacobs}, {Justen}, {Klein},
  {Mertens}, {Okada}, {Parikka}, {P{\"u}tz}, {Reyes}, {Richter}, {Ricken},
  {Riquelme}, {Rothbart}, {Schneider}, {Simon}, {Wienold}, {Wiesemeyer},
  {Ziebart}, {Fusco}, {Rosner}, \& {Wohler}}]{risacher2018}
{Risacher}, C., {G{\"u}sten}, R., {Stutzki}, J., {et~al.} 2018, Journal of
  Astronomical Instrumentation, 7, 1840014

\bibitem[{{Rohlfs} \& {Wilson}(2006)}]{rohlfs2006}
{Rohlfs}, K. \& {Wilson}, T.~L. 2006, {Tools of radio astronomy} (Tools of
  radio astronomy, 4th rev.~and enl.~ed., by K.~Rohlfs and T.L.~Wilson.~
  Berlin: Springer, 2006)

\bibitem[{{R{\"o}llig} \& {Ossenkopf}(2013)}]{roellig2013}
{R{\"o}llig}, M. \& {Ossenkopf}, V. 2013, \aap, 550, A56

\bibitem[{{Russeil} {et~al.}(2013){Russeil}, {Schneider}, {Anderson},
  {Zavagno}, {Molinari}, {Persi}, {Bontemps}, {Motte}, {Ossenkopf},
  {Andr{\'e}}, {Arzoumanian}, {Bernard}, {Deharveng}, {Didelon}, {Di
  Francesco}, {Elia}, {Hennemann}, {Hill}, {K{\"o}nyves}, {Li}, {Martin},
  {Nguyen Luong}, {Peretto}, {Pezzuto}, {Polychroni}, {Roussel}, {Rygl},
  {Spinoglio}, {Testi}, {Tig{\'e}}, {Vavrek}, {Ward-Thompson}, \&
  {White}}]{russeil2013}
{Russeil}, D., {Schneider}, N., {Anderson}, L.~D., {et~al.} 2013, \aap, 554,
  A42

\bibitem[{{Sandell} {et~al.}(2020){Sandell}, {Wright}, {G{\"u}sten},
  {Wiesemeyer}, {Reyes}, {Mookerjea}, \& {Corder}}]{sandell2020}
{Sandell}, G., {Wright}, M., {G{\"u}sten}, R., {et~al.} 2020, \apj, 904, 139

\bibitem[{{Schneider} {et~al.}(2020){Schneider}, {Simon}, {Guevara},
  {Buchbender}, {Higgins}, {Okada}, {Stutzki}, {G{\"u}sten}, {Anderson},
  {Bally}, {Beuther}, {Bonne}, {Bontemps}, {Chambers}, {Csengeri}, {Graf},
  {Gusdorf}, {Jacobs}, {Justen}, {Kabanovic}, {Karim}, {Luisi}, {Menten},
  {Mertens}, {Mookerjea}, {Ossenkopf-Okada}, {Pabst}, {Pound}, {Richter},
  {Reyes}, {Ricken}, {R{\"o}llig}, {Russeil}, {S{\'a}nchez-Monge}, {Sandell},
  {Tiwari}, {Wiesemeyer}, {Wolfire}, {Wyrowski}, {Zavagno}, \&
  {Tielens}}]{schneider2020}
{Schneider}, N., {Simon}, R., {Guevara}, C., {et~al.} 2020, \pasp, 132, 104301

\bibitem[{{Sch{\"o}ier} {et~al.}(2005){Sch{\"o}ier}, {van der Tak}, {van
  Dishoeck}, \& {Black}}]{schoeier2005}
{Sch{\"o}ier}, F.~L., {van der Tak}, F.~F.~S., {van Dishoeck}, E.~F., \&
  {Black}, J.~H. 2005, \aap, 432, 369

\bibitem[{{Scoville} {et~al.}(1986){Scoville}, {Sargent}, {Sanders},
  {Claussen}, {Masson}, {Lo}, \& {Phillips}}]{scoville1986}
{Scoville}, N.~Z., {Sargent}, A.~I., {Sanders}, D.~B., {et~al.} 1986, \apj,
  303, 416

\bibitem[{{Sharma} {et~al.}(2017){Sharma}, {Pandey}, {Ojha}, {Bhatt}, {Ogura},
  {Kobayashi}, {Yadav}, \& {Pandey}}]{sharma2017}
{Sharma}, S., {Pandey}, A.~K., {Ojha}, D.~K., {et~al.} 2017, \mnras, 467, 2943

\bibitem[{{Simon} {et~al.}(1997){Simon}, {Stutzki}, {Sternberg}, \&
  {Winnewisser}}]{simon1997b}
{Simon}, R., {Stutzki}, J., {Sternberg}, A., \& {Winnewisser}, G. 1997, \aap,
  327, L9

\bibitem[{{Spitzer}(1998)}]{spitzer1998}
{Spitzer}, L. 1998, Physical Processes in the interstellar medium (John Wiley
  and Sons, Inc.)

\bibitem[{{Stutzki} \& {Guesten}(1990)}]{stutzki1990}
{Stutzki}, J. \& {Guesten}, R. 1990, \apj, 356, 513

\bibitem[{{Syed} {et~al.}(2020){Syed}, {Wang}, {Beuther}, {Soler}, {Rugel},
  {Ott}, {Brunthaler}, {Kerp}, {Heyer}, {Klessen}, {Henning}, {Glover},
  {Goldsmith}, {Linz}, {Urquhart}, {Ragan}, {Johnston}, \& {Bigiel}}]{syed2020}
{Syed}, J., {Wang}, Y., {Beuther}, H., {et~al.} 2020, \aap, 642, A68

\bibitem[{{Tauber} {et~al.}(1995){Tauber}, {Lis}, {Keene}, {Schilke}, \&
  {Buettgenbach}}]{tauber1995}
{Tauber}, J.~A., {Lis}, D.~C., {Keene}, J., {Schilke}, P., \& {Buettgenbach},
  T.~H. 1995, \aap, 297, 567

\bibitem[{{Tauber} {et~al.}(1994){Tauber}, {Tielens}, {Meixner}, \&
  {Goldsmith}}]{tauber1994}
{Tauber}, J.~A., {Tielens}, A.~G.~G.~M., {Meixner}, M., \& {Goldsmith}, P.~F.
  1994, \apj, 422, 136

\bibitem[{{Taylor} {et~al.}(2003){Taylor}, {Gibson}, {Peracaula}, {Martin},
  {Landecker}, {Brunt}, {Dewdney}, {Dougherty}, {Gray}, {Higgs}, {Kerton},
  {Knee}, {Kothes}, {Purton}, {Uyaniker}, {Wallace}, {Willis}, \&
  {Durand}}]{taylor2003}
{Taylor}, A.~R., {Gibson}, S.~J., {Peracaula}, M., {et~al.} 2003, \aj, 125,
  3145

\bibitem[{{Tielens} {et~al.}(1993){Tielens}, {Meixner}, {van der Werf},
  {Bregman}, {Tauber}, {Stutzki}, \& {Rank}}]{tielens1993}
{Tielens}, A.~G.~G.~M., {Meixner}, M.~M., {van der Werf}, P.~P., {et~al.} 1993,
  Science, 262, 86

\bibitem[{{Tiwari} {et~al.}(2021){Tiwari}, {Karim}, {Pound}, {Wolfire},
  {Jacob}, {Buchbender}, {G{\"u}sten}, {Guevara}, {Higgins}, {Kabanovic},
  {Pabst}, {Ricken}, {Schneider}, {Simon}, {Stutzki}, \&
  {Tielens}}]{tiwari2021}
{Tiwari}, M., {Karim}, R., {Pound}, M.~W., {et~al.} 2021, \apj, 914, 117

\bibitem[{{Townsley} {et~al.}(2018){Townsley}, {Broos}, {Garmire}, {Anderson},
  {Feigelson}, {Naylor}, \& {Povich}}]{townsley2018}
{Townsley}, L.~K., {Broos}, P.~S., {Garmire}, G.~P., {et~al.} 2018, \apjs, 235,
  43

\bibitem[{{Ungerechts} {et~al.}(2000){Ungerechts}, {Umbanhowar}, \&
  {Thaddeus}}]{ungerechts2000}
{Ungerechts}, H., {Umbanhowar}, P., \& {Thaddeus}, P. 2000, \apj, 537, 221

\bibitem[{{Van Oort} {et~al.}(2019){Van Oort}, {Xu}, {Offner}, \&
  {Gutermuth}}]{vanoort2019}
{Van Oort}, C.~M., {Xu}, D., {Offner}, S. S.~R., \& {Gutermuth}, R.~A. 2019,
  \apj, 880, 83

\bibitem[{{Weaver} {et~al.}(1977){Weaver}, {McCray}, {Castor}, {Shapiro}, \&
  {Moore}}]{weaver1977}
{Weaver}, R., {McCray}, R., {Castor}, J., {Shapiro}, P., \& {Moore}, R. 1977,
  \apj, 218, 377

\bibitem[{{Werner} {et~al.}(2004){Werner}, {Roellig}, {Low}, {Rieke}, {Rieke},
  {Hoffmann}, {Young}, {Houck}, {Brandl}, {Fazio}, {Hora}, {Gehrz}, {Helou},
  {Soifer}, {Stauffer}, {Keene}, {Eisenhardt}, {Gallagher}, {Gautier}, {Irace},
  {Lawrence}, {Simmons}, {Van Cleve}, {Jura}, {Wright}, \&
  {Cruikshank}}]{werner2004}
{Werner}, M.~W., {Roellig}, T.~L., {Low}, F.~J., {et~al.} 2004, \apjs, 154, 1

\bibitem[{{Wilson} \& {Rood}(1994)}]{wilson1994}
{Wilson}, T.~L. \& {Rood}, R. 1994, \araa, 32, 191

\end{thebibliography}

\end{document}